\documentclass[showpacs,superscriptaddress,reprint,onecolumn,nofootinbib]{article}
\pdfoutput=1
\usepackage{amsmath}
\usepackage{amsbsy}
\usepackage{amssymb}
\usepackage{graphicx}
\usepackage{geometry}
\usepackage{rotating}
\usepackage[affil-it]{authblk}

\usepackage[english]{babel}
\usepackage[T1]{fontenc}
\usepackage[utf8]{inputenc}
\IfFileExists{lmodern.sty}{\usepackage{lmodern}}{}
\usepackage{color}
\definecolor{darkblue}{rgb}{0,0,0.6}
\definecolor{darkred}{rgb}{0.6,0,0}
\definecolor{darkgreen}{rgb}{0,0.6,0}
\usepackage{setspace}
\usepackage{environ}
\usepackage{wasysym}
\numberwithin{equation}{section}
\usepackage[colorlinks=true,urlcolor=darkblue,citecolor=darkblue,linkcolor=darkred,hyperindex=true,hyperfootnotes=false]{hyperref}
\usepackage{chngcntr}
\usepackage[makeroom]{cancel}
\counterwithin{figure}{section}
\usepackage{multirow}
\usepackage{enumerate}

\newcommand{\br}{\mathbf r}
\newcommand{\muint}{\mathfrak{u}_{\rm int}}

\newcommand{\eps}{\varepsilon}

\newcommand\bfJ{{\bf J}}

\usepackage{tikz}

\newcommand{\bu}{\bold{u}}

\newcommand{\bm}{\bold{m}}

\newcommand{\bbeta}{\boldsymbol{\eta}}

\newcommand\hmuint{\hat{\mathfrak{u}}_{\rm int}}

\title{How active field theories couple to external potentials}

\author[1]{Yariv Kafri}

\author[2]{Julien Tailleur}

\affil[1]{Department of Physics, Technion – Israel Institute of Technology, Haifa 32000, Israel}

\affil[2]{Department of Physics, Massachusetts Institute of Technology, Cambridge, Massachusetts 02139, USA}

\date{\today}

\usepackage[backend=biber,style=numeric,sorting=none,natbib=true,hyperref]{biblatex}
\bibliography{Pot.bib}

\newcommand\Da{D_{a}}
\newcommand\Dp{D_{p}}
\newcommand\Pe{\mathrm{Pe}}

\begin{document}

\maketitle

\begin{abstract}
We study the simplest terms that need to be included in active field theories to couple them to external potentials. 
To do so, we consider active Brownian particles and implement a systematic perturbative expansion in the particle persistence time. 
The result is a non-trivial coupling between density and potential gradients, which accounts for the nonequilibrium features of active particles in the presence of an external potential, from boundary accumulation to far-field density modulation. 
We show how the method can be applied to particles interacting via pairwise forces and to spatial modulations of the propulsion speed.
\end{abstract}

\tableofcontents

\section{Introduction}

The coupling between active particles and their environment leads to a wealth of interesting behaviours~\cite{bechinger_active_2016,granek2024colloquium}, ranging from boundary accumulation~\cite{tailleur_statistical_2008,elgeti2009self,elgeti2013wall} to ratchet effects~\cite{galajda2007wall,di2010bacterial,sokolov2010swimming} through the destruction of bulk phase transitions due to boundary disorder~\cite{dor2022disordered}.

At the microscopic scale, it is easy to account for the coupling between active dynamics and an external potential $V(\br)$. For non-interacting active Brownian particles (ABPs) in $d$ spatial dimensions, this takes the form of a Langevin equation
\begin{equation}\label{eq:micropot}
    \dot \br= v_a \bu -\mu \nabla V(\br)+\sqrt{2 \Dp} \bbeta\;,
\end{equation}
where $v_a$ is the active propulsion speed, $\bu$ is the particle orientation vector, $\mu$ is its mobility, $\Dp$ is a passive diffusivity, and $\bbeta$ is a $d$-dimensional unit-variance centred Gaussian white noise. In addition, the particle's orientation undergoes rotational diffusion on the unit  sphere at rate $\tau_r^{-1}$, leading to an active persistence time $\tau_a=\tau_r/(d-1)$ defined through $
\langle \bu(t) \cdot \bu(0) \rangle=\exp[-t/\tau_a]$. This dynamics translates into a Fokker-Planck equation for the probability density $\psi$ to find the particle at position $\br$ with direction $\bu$, which reads
\begin{equation}\label{eq:Fokker-Planck}
    \partial_t \psi(\br,\bu,t)=-\nabla \cdot [v_a \bu \psi(\br,\bu,t)-\mu \nabla V(\br) \psi(\br,\bu,t)] + \Dp \nabla^2 \psi(\br,\bu,t) +\frac 1 {\tau_r}\nabla^2_\bu \psi(\br,\bu,t)\;.
\end{equation}
Equations~\eqref{eq:micropot} and~\eqref{eq:Fokker-Planck} allow one to study active particles in external potentials at the microscopic scale~\cite{fily2014dynamics,hennes2014self,solon_active_2015,wagner2017steady,zhao2025active}, but they do not allow assessing how universal the resulting properties are.

At the other end of the spectrum, active field theories have proven powerful tools to account for the generic collective behaviours of active particles~\cite{marchetti2013hydrodynamics,cates2025active}, and their derivation have attracted a lot of effort and interest~\cite{tailleur_statistical_2008,bialke2013microscopic,solon_generalized_2018-1,speck2021coexistence,omar2023mechanical,vrugt2023derive,li2023towards,bickmann2023active,kalz2024field,soto2024kinetic,neville2025hydrodynamic,vrugt2026microscopic}.
Coupling phenomenological field theories to external potentials is, however, difficult, due to the lack of a free energy. 
At large scale, it is known that, in the absence of a potential, the dynamics~\eqref{eq:Fokker-Planck} leads to a diffusive behaviour, with diffusivity $\Dp+\Da$, where $\Da=v_a^2 \tau_a /d$ is the contribution of self-propulsion to the large scale diffusivity~\cite{cates_when_2013}. 
A coupling that would stem from an effective large-scale free energy would thus lead to a large-scale hydrodynamic equation for the probability density field $\rho(\br,t)=\int d\bu \psi (\br,\bu,t)$ given by
\begin{equation}\label{eq:equilibriumbete}
    \partial_t \rho(\br,t) = \nabla \cdot [ (\Dp+\Da)\nabla \rho(\br,t) + \mu\rho(\br,t) \nabla V(\br)]\;.
\end{equation}
This effective theory captures the equilibrium limit of active particles interacting with an external potential in the $\tau_a\to 0$ limit, holding $\Da$ finite~\cite{solon_active_2015,fodor_how_2016}. However, it fails at capturing any of their interesting nonequilibrium properties. 
A natural question is then how one should complement~\eqref{eq:equilibriumbete} to account for the leading non-equilibrium contribution due to external potentials.

In this article, we derive such an equation in a systematic way.
Introducing the active persistence length $\ell_a=v_a\tau_a$, we
show that the leading-order correction to the equilibrium dynamics~\eqref{eq:equilibriumbete} reads
\begin{eqnarray}
\partial_t\rho
&=& \nabla \cdot \big[ (\Dp+\Da)\nabla\rho
+\mu \rho \nabla V\big] \nonumber\\\label{eq:potfieldtheorytd}
&-&\ell_a^2 \tilde\gamma \Da\nabla^4\rho
+\frac{\mu \ell_a^2}{d}\nabla\otimes\nabla : \nabla V\otimes\nabla\rho -\frac{\mu \ell_a^2}{d}\nabla^2 \big[ \nabla \cdot(\rho \nabla V) \big]\;.
\end{eqnarray}
where $\tilde \gamma$ is a positive number in $d>1$, defined as
\begin{equation}
\tilde \gamma
=
\frac{4d-1}{d^2(d+2)}\;.
\end{equation}
Formally, Eq.~\eqref{eq:potfieldtheorytd} results from an expansion in a small parameter $\eps\propto\tau_a/\tau_V$, valid up to order $\eps^2$, where $\tau_V$ is the relaxation time in the potential of a noiseless overdamped dynamics.
As for the field theory of interacting scalar active matter~\cite{wittkowski_scalar_2014,nardini_entropy_2017,solon_generalized_2018}, the non-equilibrium nature of the dynamics becomes apparent at fourth order in gradients. We also note that the equation for the steady-state of dynamics~\eqref{eq:potfieldtheorytd}, to the same order in $\ell_a^2$, takes a simpler form\footnote{This simplification occurs by noting that the $\mathcal{O}(\ell_a^2)$ in Eq.~\eqref{eq:potfieldtheorytd} can be rewritten in terms of the leading order in $\ell_a^2$ of $\partial_t\rho$. This cancels the last term in Eq.~\eqref{eq:potfieldtheorytd} at the cost of renormalizing $\tilde \gamma$ into $\gamma/d$ and introducing an extra coefficient $D_p$ in the prefactor of $\nabla^4 \rho$. See Sec.~\ref{sec:derivation} for details.}:
\begin{eqnarray} \label{eq:potfieldtheoryss}
    0&=&\nabla \cdot \big[(\Dp+\Da) \nabla \rho
+\mu \rho \nabla V\big]\notag\\
&+& \frac{\ell_a^2}{d} \Big[\Dp+\gamma \Da \Big]\nabla^4\rho
+\frac{\mu\ell_a^2}{d}\nabla\otimes\nabla:\nabla V\otimes\nabla\rho\;,
\end{eqnarray}
with
\begin{equation}
    \gamma=\frac{(d-1)^2}{d(d+2)}\;.
\end{equation}

In Eqs.~\eqref{eq:potfieldtheorytd} and~\eqref{eq:potfieldtheoryss}, the terms $\propto\nabla^4 \rho$ are expected, and the real addition due to activity is the coupling between $\nabla V$ and $\nabla \rho$. 
Most interesting is the term which involves $\nabla V \otimes \nabla \rho$ and couples them in a tensorial manner. Such a term would be missed by coupling active model B or B+ to a potential by simply complementing the current with a term $-\rho\nabla V$. We note that this  term implies that currents in, say, the $\hat{x}$ direction can be generated by potentials varying in the $\hat{y}$ direction, in the presence of a density gradient along $\hat x$. 
Such fluxes have for instance recently been shown to play a crucial role in the wetting of hard surfaces by active fluids~\cite{zhao2024active}. 

Below, we first provide a simple derivation of Eqs.~\eqref{eq:potfieldtheorytd} and~\eqref{eq:potfieldtheoryss} in Sec.~\ref{sec:derivation} for $D_p=0$, and show how they predict the emergence of boundary accumulation in Sec.~\ref{sec:boundary}. The case $D_p\neq 0$ is treated in Appendix~\ref{sec:Dp}.
We then consider the response to a localized inclusion in Sec.~\ref{sec:localized}, and generalize the method to construct the hydrodynamics of ABPs interacting via pairwise forces in Sec.~\ref{sec:pairwise}. In Sec.~\ref{sec:derivationNONU}, we consider the case of non-uniform activity.
Finally, we generalize Eqs.~\eqref{eq:potfieldtheorytd} and~\eqref{eq:potfieldtheoryss} to the case of run-and-tumble particles  in Appendix~\ref{sec:RTP} .

\section{Derivation}\label{sec:derivation} 
For the sake of clarity, we present the derivation for $D_p=0$ and defer the general case to Appendix~\ref{sec:Dp}. The Fokker-Planck equation then simplifies to
\begin{equation}
\partial_t \psi(\mathbf r,\mathbf u,t)
=
-\nabla \cdot \Big[\big[v_a\mathbf u-\mu\nabla V(\mathbf r)\big] \psi(\mathbf r,\mathbf u,t)\Big]
+\frac{1}{\tau_r}\nabla_{\mathbf u}^2\psi(\mathbf r,\mathbf u,t)
\end{equation}

We work in the limit where the persistence length and time  of the active dynamics are much smaller than the length scale $\ell_V$ and the time scale $\tau_V$ associated to the external potential, respectively. To make these scales explicit, we write
\begin{equation}
    V(\mathbf r)=V_0 \hat V \left(\frac{\mathbf r}{\ell_V}\right)\;\quad\text{and}\quad
    \tau_V=\frac{\ell_V}{v_V}=\frac{\ell_V^2}{\mu V_0}\;,
\end{equation} 
where $v_{V}=\frac{\mu V_0}{\ell_V}$ is the typical drift speed induced by the potential and $\tau_V$ is the typical relaxation time of a noiseless overdamped particle in the potential. We then change variables to $\mathbf r=\ell_V \hat{\mathbf r}$ and
$t=\tau_V \hat t$. Rescaling the probability density $\psi(\mathbf r,\mathbf u,t)=\ell_V^{-d} \hat\psi(\hat{\mathbf r},\mathbf u,\hat t)$, and dropping the hat notation for clarity, we get
\begin{equation}
\partial_t \psi(\mathbf r,\mathbf u,t)
=
-\nabla\cdot\Big[\Big(\sqrt{\frac{\mathrm{Pe}}{\eps}}\mathbf u
-\nabla V(\mathbf r)\Big)\psi(\mathbf r,\mathbf u,t)\Big]
+\frac{1}{\eps}\nabla_{\mathbf u}^2\psi(\mathbf r,\mathbf u,t)\;, \label{eq:rescaledFP}
\end{equation}
where we introduced the small parameter $\eps$ and a P\'eclet number $\Pe$ through
\begin{equation}
\eps=\frac{\tau_r}{\tau_V}=\frac{(d-1)\tau_a}{\tau_V}\qquad\text{and}\qquad
\Pe = \frac{d(d-1)\Da}{ v_V \ell_V}=\frac{v_a^2\tau_r}{v_V \ell_V}\;.
\end{equation}
The P\'eclet number compares advection due to the potential to active diffusion on the potential length scale $\ell_V$. We note that the limit we consider, $\tau_r\to 0$ with $D_a$ finite, leads to $\eps\to 0$ with $\mathrm{Pe}\sim O(1)$. 

To implement a small $\eps$ expansion, we use \eqref{eq:rescaledFP} to derive equations for the density, magnetization, and nematic tensor, defined as
\begin{equation}
\rho(\mathbf r,t) = \int \psi d\mathbf u\,, \qquad
m_i(\mathbf r,t) = \int u_i \psi  d\mathbf u \,, \qquad
Q_{ij}(\mathbf r,t)
=\int
\Big(u_i u_j - \frac{\delta_{ij}}{d}\Big)\psi  d\mathbf u \;.
\end{equation}
Their evolution equations can be obtained from the Fokker-Planck equation~\eqref{eq:rescaledFP}. It is convenient to introduce the linear operators
\begin{equation}\label{eq:LinOp}
\mathcal L_m \equiv    \partial_t - \partial_j(\partial_j V \cdot)+\frac{d-1}{\eps}\qquad\text{and}\qquad \mathcal L_Q \equiv \partial_t -\partial_k(\partial_k V\cdot)+\frac{2d}{\eps}\;.
\end{equation}
In Eq.~\eqref{eq:LinOp}, implicit summation over repeated indices is assumed, a convention we use throughout the article. The dynamics of $\rho$, $m_i$ and $Q_{ij}$ then take the form
\begin{eqnarray}
    \partial_t \rho &=& -\nabla \cdot \Big(
\sqrt{\frac{\Pe}{\eps}}\mathbf m - \rho \nabla V \Big) \label{eq:denh}\\
\mathcal L_m m_i &=&-\sqrt{\frac{\Pe}{\eps}}\Big(\frac{1}{d}\partial_i \rho+\partial_j Q_{ij}\Big)\label{eq:magh} \\
\mathcal L_Q Q_{ij}&=&-\sqrt{\frac{\Pe}{\eps}}\left\{\frac{1}{d+2}\partial_k\big(\delta_{ij} m_k+\delta_{jk} m_i+\delta_{ki} m_j\big)-\frac{\delta_{ij}}{d}\partial_k m_k+\partial_k \mathcal \chi_{ijk}\right\} \;, \label{eq:Qh}
\end{eqnarray}
where we have denoted $\mathcal{\chi}$ the third spherical harmonics of $\psi$, defined as\footnote{$\chi$ is a rank $3$ tensor which is totally symmetric and traceless under contraction of any two indices.}
\begin{equation}
\mathcal \chi_{ijk}=\int u_i u_j u_k \psi d\mathbf u-\frac{1}{d+2}\left(\delta_{ij} m_k+\delta_{jk} m_i+\delta_{ki} m_j\right)\;.
\end{equation}
Since $\mathcal{L}_m$ and $\mathcal{L}_Q$ scale as $1/\eps$,  the structure of the equations suggests that, as $\eps\to 0$,
$\rho = O(1)$, $\bm = O(\sqrt{\eps})$, $Q = O(\eps)$, and $\mathcal \chi = O(\eps^{3/2})$. 
Since we are interested in the dynamics and steady-state to leading order in $\epsilon$, this legitimates truncating the description at the order of $Q$ and neglecting higher harmonics.

Equation~\eqref{eq:denh} then implies that, to obtain the dynamics of $\rho$ to $O(\eps)$, we need to determine $\bm$ to $O(\eps^{3/2})$. To do so, we invert the linear operators $\mathcal L_m$ and $\mathcal L_Q$ as
\begin{eqnarray}
\mathcal L_m^{-1}
&=&\frac{\eps}{d-1}\left[1-\frac{\eps}{d-1}\big[\partial_t-\partial_j(\partial_j V\,\cdot)\big]\right] + O(\eps^3) \label{eq:Lmm1}\\
\mathcal L_Q^{-1}&=&\frac{\eps}{2d}
+ O(\eps^2)\;.
\end{eqnarray}
We note that the leading-order contributions in $\epsilon$ to $\mathcal L_m$ and $\mathcal L_Q$ are constant, hence trivially invertible.
The inversion remains possible as long as the field variations with time and length are small enough so that both series converge. 
As usual, this must be determined self-consistently. To leading order, we then find 
\begin{equation}
Q_{ij}=-\frac{\sqrt{\Pe\,\eps}}{2d(d+2)}\left(\partial_i m_j+\partial_j m_i-\frac{2}{d}\delta_{ij}\nabla \cdot m\right)+ O(\eps^2).
\end{equation}
Using $\mathcal L_m^{-1}$ to order $\eps$ then gives $m_i=-\frac{\sqrt{\Pe\,\eps}}{d(d-1)}\partial_i \rho + O(\eps^{3/2})$, which shows that
\begin{equation}\label{eq:leadingQ}
Q_{ij}=\frac{\Pe\,\eps}{d^2(d-1)(d+2)}\left(\partial_i\partial_j \rho-\frac{1}{d}\delta_{ij}\nabla^2\rho\right)+ O(\eps^2).
\end{equation}
Using Eqs.~\eqref{eq:magh},~\eqref{eq:Lmm1}, and~\eqref{eq:leadingQ} then gives for $\bm$:
\begin{equation}\mathbf m=-\frac{\sqrt{\Pe\,\eps}}{d(d-1)}\nabla\rho -\frac{\eps^{3/2}}{d(d-1)}
\Bigg[\frac{(\Pe)^{3/2}}{d^2(d+2)}\nabla(\nabla^2\rho)-\frac{\sqrt{\Pe}}{d-1}\Big(\nabla\partial_t\rho-(\nabla^2 V)\nabla\rho-(\nabla V\!\cdot\!\nabla)\nabla\rho\Big)\Bigg]+ O(\eps^{5/2}).
\end{equation}
Finally, using this in Eq. \eqref{eq:denh} yields 
\begin{eqnarray}
    \partial_t \rho&=& \frac{\Pe}{d(d-1)}\nabla^2 \rho+\nabla \cdot (\rho \nabla V) \nonumber \\
    &+&\eps \Bigg[\frac{\Pe^2}{d^3(d-1)(d+2)}(\nabla^4\rho)-\frac{\Pe}{d(d-1)^2}\Big(\nabla^2\partial_t\rho-\nabla \otimes \nabla: \nabla V\otimes\nabla\rho\Big)\Bigg]+ O(\eps^{2}) \;.\label{eq:selcondsptrho}
\end{eqnarray}
Note the presence of $\partial_t \rho$ on the right-hand side, which is a retardation effect due to the dynamics of $\bm$. At this stage, we can already recover the steady-state solution Eq.~\eqref{eq:potfieldtheoryss} by setting $\partial_t \rho=0$, rescaling back space and time to our initial variables, and reintroducing $D_p$. 

Then, using Eq.~\eqref{eq:selcondsptrho} recursively to replace $\partial_t \rho$ in the right-hand side  by its $O(\epsilon^0)$ estimate yields
\begin{eqnarray}
    \partial_t \rho&=& \frac{\Pe}{d(d-1)}\nabla^2 \rho+\nabla \cdot (\rho \nabla V) \nonumber \\
    &+&\eps \Bigg[\frac{\Pe^2}{d^3(d-1)(d+2)}(\nabla^4\rho)+\frac{\Pe}{d(d-1)^2}\nabla \otimes \nabla: \nabla V\otimes\nabla\rho\nonumber\\
    &-&\frac{\Pe}{d(d-1)^2}\nabla^2\Big[ \frac{\Pe}{d(d-1)}\nabla^2 \rho+\nabla \cdot (\rho \nabla V) \Big]\Bigg]+ O(\eps^{2}) \;\\
    &=& \frac{\Pe}{d(d-1)}\nabla^2 \rho+\nabla \cdot (\rho \nabla V) \nonumber \\
    &+&\eps \Bigg[\frac{\Pe^2 (1-4d)}{d^3(d-1)^3(d+2)}\nabla^4\rho+\frac{\Pe}{d (d-1)^2}\nabla \otimes \nabla: \nabla V\otimes\nabla\rho\notag\\
    &-&\frac{\Pe}{d (d-1)^2}\nabla^2\nabla \cdot (\rho \nabla V)\Bigg]+ O(\eps^{2})
\end{eqnarray}
Turning back to the initial variables leads to Eq.~\eqref{eq:potfieldtheorytd} upon the restoration of $\Dp\neq 0$.

\section{Accumulation near boundaries---The case of harmonic confinement}
\label{sec:boundary}
To get a feeling for the new physics that the fourth-order gradient terms generate, compared to the effective equilibrium, we solve for the steady state in a harmonic potential $V(r)=k r^2/2$ in dimension $d>1$.

The steady state solves
\begin{equation} 
    0=\nabla \cdot \big[D \nabla \rho
+\mu \rho \nabla V\big]+\frac{\ell_a^2}{d} \tilde D \nabla^4\rho
+\frac{\ell_a^2}{d}\mu\nabla\otimes\nabla:\nabla V\otimes\nabla\rho\;,
\end{equation}
where we introduced $D=\Dp+\Da$ and $\tilde D=\Dp+ \gamma \Da$.
In spherical coordinates this takes the form
\begin{eqnarray}
0&=&D\left(\rho''+\frac{d-1}{r}\rho'\right)+\mu k\,(d\rho+r\rho')
+\frac{\ell_a^2 \tilde D}d \Big( \rho''''+\frac{2(d-1)}{r}\rho'''+\frac{(d-1)(d-3)}{r^2}\rho''-\frac{(d-1)(d-3)}{r^3}\rho' \Big) \nonumber \\
&+&\frac{\ell_a^2}{d}\mu k\left(r\rho'''+2d\,\rho''+d(d-1)\frac{\rho'}{r}\right).
\end{eqnarray}
As expected, one recovers an effective equilibrium solution for $\ell_a=0$, which reads
\begin{equation}
\rho^{(0)}(r)=\mathcal N \exp\left(-\frac{\mu}{D} V(r)\right)
=\mathcal N \exp\left(-\frac{\mu k}{2 D}\,r^2\right)\;,
\end{equation}
with $\mathcal N$ a normalization. To obtain the next order we write
\begin{equation}\label{eq:scaling}
    \rho(r)=\rho^{(0)}(r)\Big[1+\ell_a^2 f(r)\Big]+O(\ell_a^4)\;.
\end{equation}
Since derivatives of $\rho^{(0)}$ are of the form of a polynomial multiplying $\rho^{(0)}$, using symmetry and the fact that the highest derivative is of order four, we can look for $f(r)$ as a polynomial:
\begin{equation}
    f(r)=c_2 r^2 + c_4 r^4 \;.
\end{equation}
\if{After some algebra, one finds
\begin{equation}
    c_2=\left(\frac{\mu k}{D_a}\right)^2\frac{(d+1)- \gamma  (d+2)}{2d}=\left(\frac{\mu k}{D_a}\right)^2\frac{d^3+2d-1}{2d^3},\qquad
c_4=\left(\frac{\mu k}{D_a}\right)^3\frac{\gamma  -1}{4d}=-\left(\frac{\mu k}{D_a}\right)^3\frac{d^3+d^2+2d-1}{4d^3(d+2)}.    
\end{equation}}\fi
After some algebra,  one finds
\begin{eqnarray}
    c_2&=&\left(\frac{\mu k}{D}\right)^2\frac{D(d+1)- \tilde D (d+2)}{2d D}=\left(\frac{\mu k}{\Da+\Dp}\right)^2\frac{(3d-1) \Da-d\Dp}{2d^2 (\Da+\Dp)}\\
c_4&=&\left(\frac{\mu k}{D}\right)^3\frac{\tilde D -D}{4dD}=\left(\frac{\mu k}{\Da+\Dp}\right)^3\frac{1-4d }{4d^2 (d+2)} \frac{\Da}{\Da+\Dp}\;.   
\end{eqnarray}
Finally, to order $\ell_a^2$, this leads to
\begin{equation}
\rho(r)=Ne^{-\frac{\mu k}{2D}r^2}
\left[1+\ell_a^2\Big[\Big(\frac{\mu k}{D}\Big)^2\frac{(3d-1)\Da-d \Dp}{2d^2D} r^2 +\Big(\frac{\mu k}{D}\Big)^3\frac{(1-4d)\Da}{4d^2(d+2) D}r^4 \Big] \right] +O(\ell_a^4).
\end{equation}
The coefficient of $r^4$ is always negative, while, for large activity, $\Da (3d-1)>d \Dp$, so the coefficient of $r^2$ is positive. 
Activity thus leads to an inverted ``$\phi^4$'' like potential term multiplying the Gaussian, which signals accumulation at the edges.
The accumulation at boundaries, which is typical of active particles, is thus already captured by the first correction to equilibrium due to the particle persistence.

\begin{figure}
    \centering
    \includegraphics[width=0.24\linewidth]{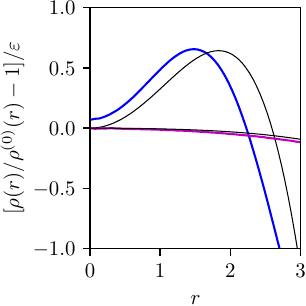}
    \includegraphics[width=0.24\linewidth]{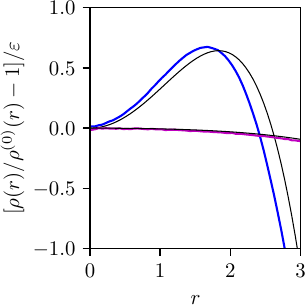}
    \includegraphics[width=0.24\linewidth]{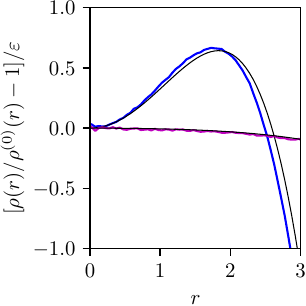}
    \includegraphics[width=0.24\linewidth]{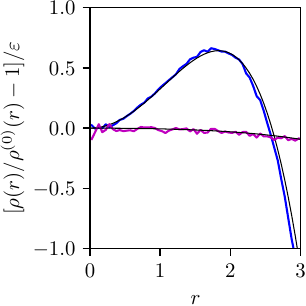}
    \caption{The probability density constructed from ABP simulations in a harmonic potentials (blue \& magenta) are shown to converge to their predicted value (black) as $\varepsilon\to 0$. Following Eq.~\eqref{eq:scaling}, the data are rescaled by their equilibrium limit $\rho^{(0)}$ and normalized by $\varepsilon\propto \ell_a^2$ to show the convergence. Accumulation away from $r=0$ develops when $(3d-1) D_a>d D_p$ (blue) and not otherwise (magenta). Parameters: $d=2$, $D_a=1$, $\mu=k=1$, $D_p=3.5$ (magenta) and $D_p=0.4$ (blue).}
    \label{fig:potential}
\end{figure}

For the perturbation theory to hold, the orientational relaxation time must be short compared to the relaxation time associated with the external force. The deterministic dynamics $\dot{\mathbf r}=-\mu k\mathbf r$ defines the relaxation time $\tau_k=(\mu k)^{-1}$. The small parameter controlling the expansion is therefore $\varepsilon=\frac{\tau_a}{\tau_k}=\tau_a \mu k \ll 1$. 
Note that, for the harmonic potential, there is no intrinsic potential length scale, but one may introduce the diffusive-confinement length $\ell_k\equiv\sqrt{\frac{D_a}{\mu k}}$. $\ell_k$ corresponds to the typical excursion of an equilibrium particle at temperature $D_a/\mu$ in the harmonic potential. On that scale, the characteristic drift velocity reads $v_k=\mu k\ell_k$. The dimensionless parameters introduced in Sec.~\ref{sec:derivation} then become $\epsilon=\frac{\tau_a}{\tau_k}=\tau_a\mu k$ and $\mathrm{Pe}=\frac{d(d-1)D_a}{v_k\ell_k}=d(d-1)$. As set in Sec.~\ref{sec:derivation},  $\mathrm{Pe}=O(1)$ while $\epsilon \ll 1$ controls the expansion, which is shown to converge on Fig.~\ref{fig:potential}.

\section{Response to a localized potential}
\label{sec:localized}

In this section we show that our perturbative theory  Eq.~\eqref{eq:potfieldtheoryss} suffices to reproduce the far-field density profile arising from a localized asymmetric potential~\cite{baek_generic_2018,granek_bodies_2020,granek2024colloquium}. For simplicity, we set $\Dp=0$ and start from
\begin{equation} \label{eq:startingloc2}
    0=\nabla \cdot \big[\Da \nabla \rho
+\mu \rho \nabla V\big]+\frac{\ell_a^2}{d} \gamma \Da \nabla^4\rho
+\frac{\ell_a^2}{d}\mu\nabla\otimes\nabla:\nabla V\otimes\nabla\rho\;\;.
\end{equation}
Since we cannot solve~\eqref{eq:startingloc2} for an arbitrary potential, we proceed perturbatively in the scale $V_0$ of the potential.

It is then useful to use $\rho=\rho_0+\phi$, with $\rho_0$ the density infinitely far away from the localized potential, and to define
\begin{equation}
    \Theta=\phi+\frac{\ell_a^2}{d}\gamma \nabla^2\phi \;.
\end{equation}
In Fourier space we then have
\begin{equation}\label{eq:chiphi}
    \phi_{\mathbf k}=\frac{\Theta_{\mathbf k}}{1-\frac{\ell_a^2}{d} \gamma k^2}\;,
\end{equation}
with Fourier conventions
\begin{equation}
    \phi_{\mathbf k}=\int d\br \phi(\br) e^{-i \mathbf k\cdot \br} \qquad\text{and}\qquad \phi(\br)= \int \frac {d {\mathbf k}} {(2\pi)^d}  \phi_{\mathbf k} e^{i \mathbf k\cdot \br} \equiv \int_{\mathbf k} \phi_{\mathbf k} e^{i \mathbf k\cdot \br}\;.
\end{equation}

In the small $k$ limit, Eq.~\eqref{eq:chiphi} gives $ \phi_{\mathbf k}\simeq \Theta_{\mathbf k}$, so that the far-field of $\Theta$ is the same as that of $\phi$.
We can then write Eq.~\eqref{eq:startingloc2} as a Poisson equation for $\Theta$:
\begin{equation}\label{eq:Dchi}
    \Da \nabla^2 \Theta=-S\;,
\end{equation}
where the source is given by
\begin{equation}\label{eq:source}
S =\mu\rho_0 \nabla^2 V +\mu \nabla \cdot(\phi\nabla V)
+\frac{\ell_a^2}{d}\mu\nabla\otimes\nabla:\nabla V\otimes\nabla\phi\;.
\end{equation}
One can then solve for $\Theta(\br)$ as
\begin{equation}
    \Theta(\mathbf r)=-\frac 1 \Da \int G(\mathbf r-\mathbf r') S(\mathbf r') d \br'\;,
\end{equation}
where $G$ is the Green's function of the Laplacian in $d$ spatial dimensions.

Expanding $\Theta$ using a multipole expansion, we find
\begin{equation}
    \Theta(\mathbf r)=M G(\mathbf r)+{\cal P}_i\partial_i G(\mathbf r) + \ldots \;,
\end{equation}
where we have introduced the monopole $M=-\Da^{-1}\int d \br' S(\mathbf r')$ and the dipole ${\cal P}_i=\Da^{-1} \int d \br' x_i S(\mathbf r')$. From the structure of $S$, one readily finds $M=0$ and
\begin{equation}\label{eq:dipoleexp}
    {\cal P}_i=-\frac \mu \Da \int d \br' \phi(\mathbf r') \partial_i V(\mathbf r')\;.
\end{equation}
We note that, due to the derivative tensor $\nabla \otimes \nabla$ in Eq.~\eqref{eq:source}, 
the $O(\ell_a^2)$ term in $S$ does not contribute to the dipole.

We now proceed to evaluate $ {\cal P}_i$ perturbatively in the strength $V_0$ of the potential, to the first non-zero order. 
We denote the $n^{\rm th}$ order by ${\cal P}_i^{(n)}$ and note that ${\cal P}_i^{(1)} \propto \int d \br' \rho_0 \partial_i V(\mathbf r')=0$.
We then Fourier-transform Eq.~\eqref{eq:startingloc2} to get:
\begin{equation} \label{eq:Kspaceeq}
   -\Da(k^2-\frac{\ell_a^2}{d} \gamma k^4)\phi_{\mathbf k} = \mu\rho_0 k^2 V_{\mathbf k}
 + \mu \int_{\mathbf q}  \mathbf k \cdot (\mathbf k-\mathbf q) \phi_{\mathbf q} V_{\mathbf k-\mathbf q}  - \frac{\ell_a^2}{d}\mu  \int_{\mathbf q}  [\mathbf k \cdot (\mathbf k-\mathbf q)] (\mathbf k \cdot \mathbf q) V_{\mathbf k-\mathbf q} \phi_{\mathbf q} \;,
\end{equation}
where we use the shorthand $\int_{\mathbf q}=\int d {\mathbf q}/(2 \pi)^d$. Since we are interested in results to order $\ell_a^2$, it is useful to introduce and expand the propagator for $\phi$ as
\begin{equation}
    \tilde G_{\mathbf k}=\frac{1}{k^2-\frac{\ell_a^2}{d} \gamma  k^4}=\frac{1}{k^2}\Big(1+\frac{\ell_a^2}{d} \gamma  k^2+O(\ell_a^4) \Big) \;.
\end{equation}
Then, using that $\phi$ vanishes when $V$ vanishes, we find to first order in the potential strength that
\begin{eqnarray}
    \phi^{(1)}_{\mathbf k}=-\frac{\mu \rho_0}{\Da} \tilde G_{\mathbf k} k^2 V_{\mathbf k} =-\frac{\mu\rho_0}{D_a} V_{\mathbf k}-\frac{\mu\rho_0 \ell_a^2\gamma}{d D_a} k^2 V_{\mathbf k}+O(\ell_a^4). \label{eq:phione}
\end{eqnarray}
We note that $\phi$ already deviates from an effective Boltzmann distribution at this order. Using Eq.~\eqref{eq:dipoleexp}, the first order correction to $\phi$ in $V_0$ then gives the second order correction to the dipole as
\begin{equation}
{\cal P}_i^{(2)}=\frac\mu\Da\int_{\mathbf k}\phi^{(1)}_{\mathbf k} i k_iV_{-\mathbf k}=0\;.
\end{equation}
The final equality holds since using Eq.~\eqref{eq:phione} shows the integrand to be odd in $k$.
Solving Eq.~\eqref{eq:Kspaceeq} iteratively and changing variables from $\mathbf q$ to $\mathbf p=\mathbf k-\mathbf q$, we find the next order correction to $\phi_{\mathbf k}$ as
\begin{eqnarray}
\phi^{(2)}_{\mathbf k}&=&\frac{\mu^2\rho_0}{k^2 D_a^2} \Bigg\{
 \int_{\mathbf p} (\mathbf k \cdot \mathbf p) V_{\mathbf p}V_{\mathbf k-\mathbf p} +\frac{\ell_a^2  \gamma }{d} \int_{\mathbf p}(\mathbf k \cdot \mathbf p) [{\mathbf k}^2+(\mathbf k-\mathbf p)^2] V_{\mathbf p} V_{\mathbf k-\mathbf p} \nonumber \\
&&-\frac{\ell_a^2}{d} \int_{\mathbf p}(\mathbf k \cdot \mathbf p)[\mathbf k \cdot (\mathbf k-\mathbf p)] V_{\mathbf p} V_{\mathbf k-\mathbf p}
\Bigg\}+O(\ell_a^4).
\end{eqnarray}
Using this in Eq.~\eqref{eq:dipoleexp}, we then find
\begin{eqnarray}
{\cal P}_i^{(3)}
&=&  \frac{\mu^3\rho_0}{D_a^3}
\int_{\mathbf p,\mathbf k}
\frac{i k_i}{\mathbf k^2}
(\mathbf k \cdot \mathbf p)\,
V_{\mathbf p} V_{\mathbf k} V_{-\mathbf k-\mathbf p} 
\nonumber \\
&+&\frac{\mu^3\rho_0 \ell_a^2 \gamma }{d D_a^3}
\int_{\mathbf p,\mathbf k}
i k_i
(\mathbf k \cdot \mathbf p)
 V_{\mathbf p} V_{\mathbf k} \,V_{-\mathbf k-\mathbf p}
\nonumber \\
&+& \frac{\mu^3\rho_0 \ell_a^2 \gamma }{d D_a^3}
\int_{\mathbf p,\mathbf k}
\frac{i k_i}{\mathbf k^2}\,
(\mathbf k \cdot \mathbf p)
(\mathbf k+\mathbf p)^2
V_{\mathbf p}V_{\mathbf k}V_{-\mathbf k-\mathbf p}
\nonumber \\
&-&\frac{\mu^3\rho_0 \ell_a^2}{D_a^3 d}
\int_{\mathbf p,\mathbf k}
\frac{i\,k_i}{\mathbf k^2}
(\mathbf k \cdot \mathbf p)
[\mathbf k \cdot (\mathbf k + \mathbf p)] 
V_{\mathbf p} V_{\mathbf k} V_{-\mathbf k-\mathbf p}   \;. 
\end{eqnarray}
The first line is $O(\ell_a^0)$ while the rest are $O(\ell_a^2)$. The integral of the $O(\ell_a^0)$ term can be written as 
\begin{equation}
 \int_{\mathbf k}\int_{\mathbf p}\int_{\mathbf q}
\delta(\mathbf k+\mathbf p+\mathbf q)
\frac{i k_i}{k^2}(\mathbf k \cdot \mathbf p) 
V_{\mathbf k} V_{\mathbf p} V_{\mathbf q}.
\end{equation}
Next, average this expression, swapping $\mathbf p$ and $\mathbf q$ and using
\begin{equation}
    \frac{1}{2}\frac{\mathbf k \cdot \mathbf p+\mathbf k \cdot \mathbf q}{\mathbf k^2}
=\frac{1}{2}\frac{\mathbf k \cdot (-\mathbf k)}{\mathbf k^2}=-\frac{1}{2} \;,
\end{equation}
where the first equality stems from the delta function, to show the $O(\ell_a^0)$ term to vanish, since
\begin{equation}
\int_{\mathbf k}\int_{\mathbf p}\int_{\mathbf q}\delta(\mathbf k+\mathbf p+\mathbf q) 
i k_i
V_{\mathbf k} V_{\mathbf p} V_{\mathbf q}
= \int d \br [\partial_i V(\br)]V^2(\br)=0 \;. 
\end{equation}
For the rest of the terms there is no a priori reason for the integrals to vanish and we finally find, in real space,
\begin{eqnarray}
    {\cal P}_i^{(3)} &=&-\frac{\mu^3\rho_0 \ell_a^2 \gamma}{2 d D_a^3} \int d {\bf r} V \partial_i (\nabla V)^2 \nonumber \\
    &-&\frac{\mu^3\rho_0 \ell_a^2\gamma }{d D_a^3} \int d {\bf r} \big[\partial_i\partial_j(\nabla^2)^{-1}V\big] (\partial_j V)(\nabla^2 V) \nonumber \\
&-&\frac{\mu^3\rho_0 \ell_a^2}{D_a^3 d} \int d {\bf r} 
\big[\partial_i\partial_j\partial_\ell(\nabla^2)^{-1}V\big]
(\partial_j V)(\partial_\ell V) \;, \label{eq:dipolefinal}
\end{eqnarray}
where $(\nabla^2)^{-1}$ is the inverse Laplacian and we omitted the argument of $V(\br)$ for clarity.

There are a few things to note. 
First, in the passive limit $\ell_a \to 0$, $\mathcal{P}$ vanishes, as expected.
Then, there are two contributions that are non-local: the $(\nabla^2)^{-1}$ is a non-local operator, which shows that the contribution to $\mathcal{P}$ in $\br$ depends on the values of the potential at both $\br$ and $\br'\neq \br$.  
Such non-local contributions are typical of non-equilibrium and active systems~\cite{granek2024colloquium}.
We note that these contributions stem both from the $\nabla^4 \rho$ (terms $\propto \gamma$ in Eq.~\eqref{eq:dipolefinal}) and from the tensorial contribution $\nabla V \otimes \nabla \rho$ (term not $\propto \gamma$ in Eq.~\eqref{eq:dipolefinal}).
As common in active systems~\cite{baek_generic_2018}, the leading order contribution appears at third order in the potential strength.

Combining everything we find that the far-field response to a localized potential is given by
\begin{equation}
    \phi \sim \frac{1}{\Omega_d}\frac{{\cal \mathbf P}^{(3)}\cdot {\bf r}}{ r^d} \;,
\end{equation}
with $\Omega_d$ the solid angle in $d$ dimensions and ${\cal \mathbf P}^{(3)}$ given by Eq.~\eqref{eq:dipolefinal}.

\if{
\section{Response to potential disorder}
\label{sec:disorder}

\subsection{Perturbative Solution}
Also here we consider the case $D_p=0$, Namely, we solve the problem
\begin{equation} 
    0=\nabla \cdot \big[\Da \nabla \rho
+\mu \rho \nabla V\big]+\frac{\ell_a^2}{d} \gamma \Da \nabla^4\rho
+\frac{\ell_a^2}{d}\mu\nabla\otimes\nabla:\nabla V\otimes\nabla\rho\;,
\end{equation}
and as before define $\rho=\rho_0+\phi$. In addition, to model a disordered potential we take in Fourier space 
\begin{equation}
\overline{V_{\mathbf k}}=0,\qquad \overline{V_{\mathbf k}V_{\mathbf k'}}=(2\pi)^d \sigma^2 \delta(\mathbf k+\mathbf k'),
\end{equation}
where the overline denotes a disorder average. The static structure factor $S(\mathbf k)$ is then
\begin{equation}
\overline{\phi_{\mathbf k}\phi_{\mathbf k'}}=(2\pi)^d\delta(\mathbf k+\mathbf k') S(\mathbf k),\qquad S(\mathbf k)=S^{(2)}(\mathbf k)+S^{(4)}(\mathbf k)+S^{(6)}(\mathbf k)+\cdots,
\end{equation}
with $S^{(n)}=O(\sigma^n)$. Since we work in Fourier space we use the approach of the previous section. Using $f(\mathbf k)=\int d^d r e^{-i{\mathbf k}\cdot {\mathbf r}}f({\mathbf r})$, and the Green's function
\begin{equation}
    G_{\mathbf k}=\frac{1}{D_a k^2-\frac{\ell^2}{d} \gamma D_a k^4}
\end{equation}
we have
\begin{eqnarray}
\phi_{\mathbf k}&=&-G_{\mathbf k} \mu \rho_0 k^2 V_{\mathbf k} \nonumber \\
&+& \mu G_{\mathbf k} \int\frac{d^d q}{(2\pi)^d}
\Big[- \mathbf k  \cdot (\mathbf k-\mathbf q)+\ell^2\,(\mathbf k \cdot (\mathbf k-\mathbf q))(\mathbf k \cdot \mathbf q)\Big] V_{\mathbf k-\mathbf q} \phi_{\mathbf q}.
\end{eqnarray}
which for brevity we write as

In what follows it is useful to write \eqref{eq:Kspaceeq} as
\begin{equation}
    \phi_{\mathbf k}=
A(\mathbf k) V_{\mathbf k}
+ \int_{\mathbf q} B(\mathbf k,\mathbf q)\,V_{\mathbf k-\mathbf q}\phi_{\mathbf q} \;.
\end{equation}
Solving to third order in the potential we have
 \begin{eqnarray}
\phi^{(1)}_{\mathbf k}&=& A(\mathbf k) V_{\mathbf k},\\
\phi^{(2)}_{\mathbf k}&=&\int_{\mathbf q} B(\mathbf k;\mathbf q) A(\mathbf q)
V_{\mathbf k-\mathbf q} V_{\mathbf q}\;,\\
\phi^{(3)}_{\mathbf k}&=&\int_{\mathbf q,\mathbf p} B(\mathbf k;\mathbf q) B(\mathbf q;\mathbf p) A(\mathbf p)V_{\mathbf k-\mathbf q} V_{\mathbf q-\mathbf p}V_{\mathbf p} \;.
\end{eqnarray}
Note that for any finite $q$ in the small $k$-limit 
\begin{equation}
 A(k)=-\frac{\mu \rho_0}{D_a-\frac{\ell_a^2}{d} \gamma D_ak^2}\simeq - \frac{\mu \rho_0}{D_a} +O(k^2) \;, 
\end{equation}
and 
\begin{eqnarray}
    B(\mathbf k ; \mathbf q)&=& \nonumber \\
    \mu G_{\mathbf k}\Big[- \mathbf k  \cdot (\mathbf k-\mathbf q)+\ell^2\,(\mathbf k \cdot (\mathbf k-\mathbf q))(\mathbf k \cdot \mathbf q)\Big] &\simeq& \frac{\mu}{D_a} \frac{\mathbf k \cdot \mathbf q}{k^2}- \frac{\mu}{D_a}- \frac{\mu\ell^2}{D_a} \frac{(\mathbf k \cdot \mathbf q)^2}{k^2}+\mathcal O(k)
\end{eqnarray}
so that the leading divergence behaves as $k^{-1}$.

\subsection{Structure Factor}
We now consider the structure factor order by order in $\sigma^2$. 
\subsubsection{Order $\sigma^2$}
Here we find
\begin{equation}
S^{(2)}(\mathbf k)=\overline{\phi^{(1)}_{\mathbf k}\phi^{(1)}_{-\mathbf k}}
=\sigma^2 A(\mathbf k)A(-\mathbf k)
=\sigma^2\frac{(\mu\rho_0)^2}{(D_a-\frac{\ell_a^2}{d} \gamma D_ak^2)^2} \;,
\end{equation}
which is {\it local}.
\subsubsection{Order $\sigma^4$}
Here we have to consider two contributions. One from $\overline{\phi^{(2)}_{\mathbf k}\phi^{(2)}_{-\mathbf k}}$ and the mixed order $\overline{\phi^{(1)}_{\mathbf k}\phi^{(3)}_{-\mathbf k}}+\overline{\phi^{(3)}_{\mathbf k}\phi^{(1)}_{-\mathbf k}}$. The latter simplifies to $2\overline{\phi^{(1)}_{\mathbf k}\phi^{(3)}_{-\mathbf k}}$
Consider first the $22$ contribution
\begin{equation}
\phi^{(2)}_{\mathbf k}\phi^{(2)}_{-\mathbf k}=\int_{\mathbf q} \int_{\mathbf Q}B(\mathbf k;\mathbf q)A(\mathbf q)B(-\mathbf k;\mathbf Q)A(\mathbf Q) V_{\mathbf k-\mathbf q}V_{\mathbf q} V_{-\mathbf k-\mathbf Q}V_{\mathbf Q}.
\end{equation}
Performing a disorder average and considering the two relevant contractions we have
\begin{eqnarray}
    \overline{\phi^{(2)}_{\mathbf k}\phi^{(2)}_{-\mathbf k}}
&=&\sigma^4\int_{\mathbf q} \Big[
B(\mathbf k;\mathbf q)A(\mathbf q)B(-\mathbf k;-\mathbf q)A(-\mathbf q)
+B(\mathbf k;\mathbf q)A(\mathbf q)B(-\mathbf k;\mathbf q-\mathbf k)A(\mathbf q-\mathbf k)
\Big] \nonumber \\
&=& \sigma^4\int_{\mathbf q} \Big[
|B(\mathbf k;\mathbf q)A(\mathbf q)|^2
+B(\mathbf k;\mathbf q)A(\mathbf q)B(-\mathbf k;\mathbf q-\mathbf k)A(\mathbf q-\mathbf k) \Big]
\end{eqnarray}
Our interest is in the small $k$ behavior. Note that $A(\mathbf q - \mathbf k) \simeq A(\mathbf q)+O(k)$. Using this we write for the  first term in the integrand
\begin{equation}
A(\mathbf q)^2 \Big[-\frac{\mu}{D_a}\frac{\mathbf k \cdot \mathbf q}{k^2} + \frac{\mu}{D_a}+O(k^0)\Big]^2
= A(\mathbf q)^2\Big[\Big(\frac{\mu}{D_a}\Big)^2 (\frac{\mathbf k \cdot \mathbf q}{k^2})^2 - 2\Big(\frac{\mu}{D_a}\Big)^2 (\frac{\mathbf k \cdot \mathbf q}{k^2}) + \Big(\frac{\mu}{D_a}\Big)^2 + \cdots\Big], \label{eq:phi2phi2}
\end{equation}
and for the second integrand
\begin{equation}
A(\mathbf q)^2 \Big[-\frac{\mu}{D_a}(\frac{\mathbf k \cdot \mathbf q}{k^2})+ \frac{\mu}{D_a}+O(k^0)\Big]\Big[+\frac{\mu}{D_a}(\frac{\mathbf k \cdot \mathbf q}{k^2}) + O(k^0)\Big]=A(\mathbf q)^2\,\Big[-\Big(\tfrac{\mu}{D_a}\Big)^2 (\frac{\mathbf k \cdot \mathbf q}{k^2})^2 + \Big(\frac{\mu}{D_a}\Big)^2 (\frac{\mathbf k \cdot \mathbf q}{k^2}) + \cdots\Big].
\end{equation}
This implies that in Eq. \ref{eq:phi2phi2} the $1/k^2$ terms cancel in the small $k$ limit. The next order is odd in $\mathbf q$ and therefore vanished. The leading order is not singular in the small $k$ limit.

Next consider 
\begin{equation}
   \overline{\phi^{(1)}_{\mathbf k}\phi^{(3)}_{-\mathbf k}} =\int_{\mathbf q,\mathbf p} B(-\mathbf k;\mathbf q) B(\mathbf q;\mathbf p) A(\mathbf p)A(\mathbf k )\overline{V_{-\mathbf k-\mathbf q} V_{\mathbf q-\mathbf p}V_{\mathbf p} V_{\mathbf k}}
\end{equation}
There are three pairings. 

\noindent {$\overline{V_{-\mathbf k - \mathbf q}V_{\mathbf k}}$ gives $\delta(\mathbf q)$. Then the second pairing gives $\delta(\mathbf p)$. Using
\begin{equation}
    B(-\mathbf k;\mathbf 0)\xrightarrow[k\to0]{}-\frac{\mu}{D_a},
\end{equation}
and to leading order $B(\mathbf 0;\mathbf p)= \lim_{\mathbf q\to 0}  \frac{\mu}{D_a} \frac{p \cdot q}{q^2}$ which is odd in $\mathbf p$ while $A(\mathbf u)$ regular as $\mathbf u=0$ gives that this contraction to leading order is zero.

\noindent {$\overline{V_{\mathbf k}V_{\mathbf q - \mathbf p}}$ gives $\mathbf p=\mathbf k + \mathbf q$. The matching contraction $\overline{V_{-\mathbf k-\mathbf q}V_{\mathbf p}}$ matches. This gives 
\begin{equation}
\sigma^4 A(\mathbf k) \int_{\mathbf q} B(-\mathbf k;\mathbf q) B(\mathbf q;\mathbf k+\mathbf q) A(\mathbf k+\mathbf q) \;.
\end{equation}
Noting that $B(-\mathbf k;\mathbf q)=+\frac{\mu}{D_a}\,\frac{\mathbf k \cdot\mathbf q}{k^2}+O(k^0)$ that $B(\mathbf q;\mathbf k+\mathbf q)=O(k)$ and $A(\mathbf k+\mathbf q)=A(\mathbf q)+O(k)$ gives for the integrand
\begin{equation}
    \left[\frac{(\mathbf k\cdot\mathbf q)}{k^2}\right]\times O(k)\times A(\mathbf q)
\end{equation}
whose angular integral vanishes. Therefore, this contractions leading contribution is finite.

\noindent {$\overline{V_{\mathbf k}V_{\mathbf p}}$. Here both contractions give $\mathbf p=-\mathbf k$. This gives
\begin{equation}
    \sigma^4 A(\mathbf k) \int_{\mathbf q} B(-\mathbf k;\mathbf q) B(\mathbf q;-\mathbf k) A(-\mathbf k) \;.
\end{equation}
Note that $A(\mathbf k)=A(-\mathbf k)$ is finite in the small $k$ limit. Also $B(\mathbf q;-\mathbf k)=- \mu \frac{q^2}{D_a q^2-\frac{\ell_a^2}{d} \gamma D_a q^4}+O(k)$. The only singular term is $B(-\mathbf k;\mathbf q)\sim +(\mu/D_a)(\mathbf k \cdot \mathbf q)/k^2$. The leading-order singular term of the integral is therefore odd in $\mathbf k$ and vanishes so that the leading order contribution is finite.

\subsubsection{Order $\sigma^6$}
Writing $\phi=\sum_{n \geq 1} \phi^{(n)}$ to this order we need to consider terms of the form $\overline{\phi^{(3)}\phi^{(3)}}$, $\overline{\phi^{(4)}\phi^{(2)}}$ and $\overline{\phi^{(5)}\phi^{(1)}}$. To proceed, it is useful to recall that the most singular behavior comes from 
\begin{equation}
 B(\mathbf k ; \mathbf q) \simeq  \frac{\mu}{D_a} \frac{\mathbf k \cdot \mathbf q}{k^2}  +O(k^0)
\end{equation}
We go over each of the cases.
\subsubsubsection{$\overline{\phi^{(5)}\phi^{(1)}}$ Contractions}
Use
\begin{equation}
\phi^{(1)}_{\mathbf k}= A(\mathbf k) V_{\mathbf k}
\end{equation}
and 
\begin{equation}
    \phi^{(5)}_{\mathbf k'}=\int_{q_1,q_2,q_3,q_4} B(\mathbf k';\mathbf q_1)B(\mathbf q_1 ; \mathbf q_2)B(\mathbf q_2 ; \mathbf q_3)B(\mathbf q_3 ; \mathbf q_4)A(\mathbf q_4)V_{\mathbf k'-\mathbf q_1}V_{\mathbf q_1-\mathbf q_2}V_{\mathbf q_2-\mathbf q_3}V_{\mathbf q_3-\mathbf q_4}V_{\mathbf q_4}
\end{equation}
Consider two representative contractions. First, $V_{\mathbf k}$ with $V_{\mathbf k'-\mathbf q_1}$. Then overall momentum conservation imposes $\mathbf q_1=0$ and the potential singular contribution from $B(\mathbf k' ; \mathbf q_1)$ vanishes. The second case is $V_{\mathbf k}$ with $V_{\mathbf q_1-\mathbf q_2}$. Then $\mathbf q_1=\mathbf k+\mathbf q_2$ so that 
\begin{equation}
    B(-\mathbf k ; \mathbf q_1)= B(-\mathbf k ; \mathbf k + \mathbf q_2)= \frac{\mu}{D_a} \frac{-\mathbf k \cdot (\mathbf k+\mathbf q_2)}{k^2}+O(k^0) \;.
\end{equation}
The leading order singular behavior then vanishes once the angle between $\mathbf k$ and $\mathbf q_2$ is integrated over.A similar behavior is obtained for all the other contractions. 

\subsubsubsection{$\overline{\phi^{(4)}\phi^{(2)}}$ Contractions}
Here we use
\begin{eqnarray}
    \phi^{(2)}_{\mathbf k}&=&\int_{\mathbf q} B(\mathbf k;\mathbf q) A(\mathbf q)
V_{\mathbf k-\mathbf q} V_{\mathbf q} \nonumber \\
 \phi^{(4)}_{\mathbf k'}&=&\int_{q_1,q_2,q_3} B(\mathbf k';\mathbf q_1)B(\mathbf q_1 ; \mathbf q_2)B(\mathbf q_2 ; \mathbf q_3)A(\mathbf q_3)V_{\mathbf k'-\mathbf q_1}V_{\mathbf q_1-\mathbf q_2}V_{\mathbf q_2-\mathbf q_3}V_{\mathbf q_3} \;.
\end{eqnarray}
Two cases have to be considered. First, $V_{\mathbf k-\mathbf q}$ and $V_{\mathbf k' - \mathbf q_1}$ which using momentum conservation gives $\mathbf q=-\mathbf q_1$ which gives a pair
$B(\mathbf k ; \mathbf p)B(\mathbf -k ; \mathbf -p)$ which gives an overall singular term of the form $\frac{\mu^2}{D_a^2} \big(\frac{\mathbf k \cdot \mathbf p}{k^2}\big)^2$. The result of the contractions yields $\mathbf k$ independent contributions. The other contraction to consider is of the type $V_{\mathbf p}V_{\mathbf k'-\mathbf q_1}$. This yields $B(\mathbf k ; \mathbf p)B(\mathbf -k ; \mathbf p - \mathbf k)$ which gives a contribution $-\frac{\mu^2}{D_a^2} \big(\frac{\mathbf k \cdot \mathbf p}{k^2}\big)^2$ and an term which after integration over the angle between $\mathbf k$ and $\mathbf p$ is zero. Since the number of pairins of each of the cases is identical and equal to three the contributions cancel and there is no $1/k^2$ contribution at this order. 

\subsubsection{$\overline{\phi^{(3)}\phi^{(3)}}$ Contractions}
Here we use
\begin{eqnarray}
    \phi^{(3)}_{\mathbf k}&=&\int_{\mathbf q_1,\mathbf p_1} B(\mathbf k;\mathbf q_1) B(\mathbf q_1;\mathbf p_1) A(\mathbf p_1)V_{\mathbf k-\mathbf q_1} V_{\mathbf q_1-\mathbf p_1}V_{\mathbf p_1} \nonumber \\
 \phi^{(3)}_{\mathbf k'}&=&\int_{\mathbf q_2,\mathbf p_2} B(\mathbf k';\mathbf q_2) B(\mathbf q_2;\mathbf p_2) A(\mathbf p_2)V_{\mathbf k'-\mathbf q_2} V_{\mathbf q_2-\mathbf p_2}V_{\mathbf p_2} \;.
\end{eqnarray}
Since this contraction yields the $1/k^2$ singular behavior we go over the contractions in more detail.

\begin{itemize}
    \item The ladder contractions $V_{\mathbf k - \mathbf q_1}V_{\mathbf k'-\mathbf q_2}$, $V_{\mathbf q_1 - \mathbf p_1}V_{\mathbf q_2-\mathbf p_2}$ and $V_{\mathbf p_1}V_{\mathbf p_2}$. Using translational invariance these give respectively $\mathbf q_1=-\mathbf q_2$, $\mathbf p_1=-\mathbf p_2$ and the overall $\delta(\mathbf k+\mathbf k')$. It useful to define
    \begin{equation}
        J_{\mathrm{lad}}(\mathbf q_1) = \int_{\mathbf p_1} A(\mathbf p_1) A(-\mathbf p_1)B(\mathbf q_1; \mathbf p_1) B(-\mathbf q_1; -\mathbf p_1).
    \end{equation}
    to write
    \begin{equation}
        S_{\mathrm{lad}}=\sigma^6 \int_{\mathbf q_1}B(\mathbf k ; \mathbf q_1)B(-\mathbf k ; -\mathbf q_1) J_{\mathrm{lad}}(\mathbf q_1)
    \end{equation}
    \item The cross ladder contractions $V_{\mathbf k - \mathbf q_1}V_{\mathbf k'-\mathbf q_2}$, $V_{\mathbf q_1 - \mathbf p_1}V_{\mathbf p_2}$ and $V_{\mathbf p_1}V_{\mathbf q_2-\mathbf p_2}$. Using translational invariance these give respectively $\mathbf p_2=\mathbf p_1-\mathbf q_1$, $\mathbf q_1=-\mathbf q_2$ and the overall $\delta(\mathbf k+\mathbf k')$. \begin{equation}
        J_{\mathrm{cross}}(\mathbf q_1) = \int_{\mathbf p_1} A(\mathbf p_1) A(\mathbf p_1 -\mathbf q_1)B(\mathbf q_1; \mathbf p_1) B(-\mathbf q_1; \mathbf p_1-\mathbf q_1).
    \end{equation}
    to write
    \begin{equation}
        S_{\mathrm{cross}}=\sigma^6 \int_{\mathbf q_1}B(\mathbf k ; \mathbf q_1)B(-\mathbf k ; -\mathbf q_1) J_{\mathrm{cross}}(\mathbf q_1)
    \end{equation}
    \item The rest of the terms can be inspected and do not yield singular contributions in the $k \to 0 $ limit.
\end{itemize}
Collecting the ladder and cross-ladder contribution gives
\begin{eqnarray}
    \overline{\phi_{\mathbf k}\phi_{-\mathbf k}}&=& \sigma^6 \int_{\mathbf q_1}B(\mathbf k ; \mathbf q_1)B(-\mathbf k ; -\mathbf q_1) \big[ J_{\mathrm{lad}}(\mathbf q_1)+ J_{\mathrm{cross}}(\mathbf q_1)] \nonumber \\
    &=&\frac{\sigma^6 \mu^2}{D_a^2} \int_{\mathbf q_1} \frac{(\mathbf k \cdot \mathbf q_1)^2}{k^4}  \big[ J_{\mathrm{lad}}(\mathbf q_1)+ J_{\mathrm{cross}}(\mathbf q_1)] +O(k^0)
\end{eqnarray}
Note that in the integral, the correction to $B$ is $O(k^{-1})$. One can check that this contribution drops out due to the integration over the angle between $\mathbf q_1$ and $\mathbf k$. 

Next, note that $A(\mathbf R \mathbf p)=A(\mathbf p)$ with $\mathbf R$ a rotation matrix. Similarly, $B(\mathbf R \mathbf p ; \mathbf R \mathbf q)= B(\mathbf p ; \mathbf q)$. Therefore, $J_{\mathrm{lad}}(\mathbf R \mathbf q)=J_{\mathrm{lad}}(\mathbf q)$  and  $J_{\mathrm{cross}}(\mathbf R \mathbf q)=J_{\mathrm{cross}}(\mathbf q)$, so that $J_{\mathrm{lad}}(\mathbf q)=J_{\mathrm{lad}}(q^2)$ and $J_{\mathrm{cross}}(\mathbf q)=J_{\mathrm{cross}}(q^2)$. Now, use
\begin{equation}
    \int d\Omega_{\hat q} \frac{(\mathbf k \cdot \mathbf q)^2}{k^4} J(q^2)=J(q^2)\frac{k_ik_j}{k^4}  \int d\Omega_{\hat q} q_i q_j=J(q^2)\frac{k_ik_j}{k^4}  \frac{\Omega_d}{d}q^2 \delta_{ij}=J(q^2)\frac{\Omega_d}{d} \frac{q^2}{k^2}
\end{equation}
where $\Omega_d=\frac{2\pi^{d/2}}{\Gamma(d/2)}$. Then we finally obtain
\begin{eqnarray}
    \overline{\phi_{\mathbf k}\phi_{-\mathbf k}}&=&\frac{\sigma^6 \mu^2}{D_a^2} \frac{\Omega_d}{(2\pi)^d} \frac{1}{dk^2}\int dq  q^{d+1} \big[J_{\mathrm{lad}}(q)+J_{\mathrm{cross}}(q) \big] +O(k^0)
\end{eqnarray}
An inspection of the integrals shows that they depend on the UV cutoff of the noise spectrum.
}\fi

\section{Pairwise interacting particles}
\label{sec:pairwise}
Let us now show how our perturbation theory allows us to build a  scalar active field theory for particles interacting through a pairwise potential. 
Since our perturbation theory works in the limit where the potential range is much larger than the particle persistence length, this allows us to estimate how non-gradient terms enter when departing from the $\tau_a\to 0$ effective equilibrium limit.
As such, this route works in a different regime than other, more phenomenological constructions~\cite{solon_generalized_2018,omar2023mechanical,speck2021coexistence}, whose goals are to describe the large-persistence regime where Motility-Induced Phase Separation~\cite{cates_motility-induced_2015} is observed. 

We consider the case where $D_p=0$, which leads to a particle dynamics given by
\begin{equation}
\dot{\br}_i = v_a\mathbf u_i - \mu \sum_j \nabla w(|\br_i-\br_j|)\;,
\end{equation}
where $w$ is a smooth, isotropic interaction potential.

Introducing the empirical measure $\hat \psi(\br,\bu)=\sum_i \delta(\br-\br_i) \delta(\bu-\bu_i)$ and density $\hat \rho(\br)=\sum_i \delta(\br-\br_i) $, we use standard methods~\cite{solon_generalized_2018} to write the time evolution of $\psi=\langle \hat \psi\rangle $ as
\begin{equation}
    \partial_t \psi(\br,\bu,t) = -\nabla\cdot\left[v_a\mathbf u \psi(\br,\bu,t)-\mu\int d\br' \nabla w(|\br-\br'|) \langle \hat \rho(\br',t)\hat\psi(\br,\bu,t)\rangle  \right]+\frac{1}{\tau_r}\nabla_{u}^2\psi\;.\label{eq:dynpsi}
\end{equation}
Introducing the excess `chemical potential'
\begin{equation}
\hmuint[\hat \rho](\br)=\int d\br'\,w(|\br-\br'|)\hat \rho(\br')\;,
\end{equation}
the evolution of $\psi$ can be concisely rewritten as
\begin{equation}
    \partial_t \psi = -\nabla\cdot\left[v_a\mathbf u \psi-\mu \langle \nabla \hmuint \hat\psi\rangle  \right]+\frac{1}{\tau_r}\nabla_{u}^2\psi\;.\label{eq:dynpsi2}
\end{equation}

\subsection{Systematic expansion}
We now use the space and time rescaling introduced in Sec.~\ref{sec:derivation}, where the scales of the interaction potential now replace those of the external potential. It is also convenient to introduce slightly different linear operators:
\begin{equation}
\mathcal L_m \equiv    \partial_t +\frac{d-1}{\eps}\qquad\text{and}\qquad \mathcal L_Q \equiv \partial_t +\frac{2d}{\eps}\;.
\end{equation}
Then, the dynamics of $\rho$, $m_i$ and $Q_{ij}$ take the forms
\begin{eqnarray}
    \partial_t \rho &=& -\nabla \cdot \Big(
\sqrt{\frac{\Pe}{\eps}}\mathbf m - \langle \hat \rho \nabla \hmuint \rangle \Big) \label{eq:denh2}\\
\mathcal L_m m_i &=&-\sqrt{\frac{\Pe}{\eps}}\Big(\frac{1}{d}\partial_i \rho+\partial_j Q_{ij}\Big)+\partial_j \langle \hat m_i \partial_j \hmuint \rangle\label{eq:magh2} \\
\mathcal L_Q Q_{ij}&=&-\sqrt{\frac{\Pe}{\eps}}\left\{\frac{1}{d+2}\partial_k\big(\delta_{ij} m_k+\delta_{jk} m_i+\delta_{ki} m_j\big)-\frac{\delta_{ij}}{d}\partial_k m_k+\partial_k \mathcal \chi_{ijk}\right\} +\partial_k \langle \hat Q_{ij} \partial_k \hmuint \rangle \;, \label{eq:Qh2}
\end{eqnarray}
where we have introduced the empirical fields 
\begin{equation}
    \hat {\mathbf m} = \sum_i \bu_i \delta(\br-\br_i)
    \qquad
    \text{and}
    \qquad \hat {\mathbf Q} = \sum_i \Big(\bu_i \otimes \bu_i - \frac{\mathbb{I}_d}d\Big) \delta(\br-\br_i)\;.
\end{equation}
and $\mathcal{\chi}$ is, again, the third spherical harmonics of $\psi$. Since $\mathcal{L}_m$ and $\mathcal{L}_Q$ scale as $1/\eps$,  the structure of the equations still suggests that $\rho = O(1)$, $\bm = O(\sqrt{\eps})$, $Q = O(\eps)$, while $\mathcal \chi = O(\eps^{3/2})$.

Equation~\eqref{eq:denh2} then implies that, to obtain the dynamics of $\rho$ to $O(\eps)$, we need to determine $\bm$ to $O(\eps^{3/2})$. To do so, we proceed as before and invert the linear operators $\mathcal L_m$ and $\mathcal L_Q$ as
\begin{eqnarray}
\mathcal L_m^{-1}
&=&\frac{\eps}{d-1}\left[1-\frac{\eps}{d-1} \partial_t\right] + O(\eps^3) \label{eq:Lmm1v}\\
\mathcal L_Q^{-1}&=&\frac{\eps}{2d}
+ O(\eps^2)\;.
\end{eqnarray}
Therefore, since $\mathbf Q$ scales as $O(\eps)$, we find
\begin{equation}
Q_{ij}=-\frac{\sqrt{\Pe\,\eps}}{2d(d+2)}\left(\partial_i m_j+\partial_j m_i-\frac{2}{d}\delta_{ij}\nabla \cdot m\right)+ O(\eps^2)\;.
\end{equation}
Using $\mathcal L_m^{-1}$ to order $\eps$ still gives $m_i=-\frac{\sqrt{\Pe\,\eps}}{d(d-1)}\partial_i \rho + O(\eps^{3/2})$, which shows that 
\begin{equation}\label{eq:leadingQv}
Q_{ij}=\frac{\Pe\,\eps}{d^2(d-1)(d+2)}\left(\partial_i\partial_j \rho-\frac{1}{d}\delta_{ij}\nabla^2\rho\right)+ O(\eps^2).
\end{equation}
We note that $\mathbf{Q}$ is unchanged from the non-interacting case. Next, using Eqs.~\eqref{eq:magh2},~\eqref{eq:Lmm1v}, and~\eqref{eq:leadingQv}  gives for $\bm$:
\begin{equation}
 m_i=-\frac{\sqrt{\Pe\,\eps}}{d(d-1)}\partial_i \rho -\frac{\eps^{3/2}}{d(d-1)}
\Bigg[\frac{(\Pe)^{3/2}}{d^2(d+2)}\partial_i(\nabla^2 \rho)-\frac{\sqrt{\Pe}}{d-1}\partial_i \partial_t\rho\Bigg]+ \frac{\eps}{d-1} \partial_j  \langle \hat{{m}}_i \otimes \partial_j \hmuint\rangle + O(\eps^{5/2}).\label{eq:mPFAPs}
\end{equation}
Finally, using this in Eq.~\eqref{eq:denh2} gives 
\begin{eqnarray}
    \partial_t \rho&=& \frac{\Pe}{d(d-1)}\nabla^2 \rho+\nabla \cdot (\langle {\hat \rho} \nabla \hmuint \rangle) \nonumber \\
    &+&\eps \Bigg[\frac{\Pe^2}{d^3(d-1)(d+2)}(\nabla^4\rho)-\frac{\Pe}{d(d-1)^2}\nabla^2\partial_t\rho\Bigg]
    -\frac{\sqrt{\Pe\,\eps}}{d-1} \nabla \otimes \nabla : \langle \hat{\mathbf{m}} \otimes \nabla \hmuint\rangle + O(\eps^{2}) \;,
\end{eqnarray}
where we note that an implicit dependence on $\eps$ remains in $\langle \hat{\mathbf{m}} \otimes \nabla \hmuint\rangle$. 
This equation can already be used in the steady state. To get a self-consistent dynamics, we use it iteratively to write
\begin{eqnarray}
    \partial_t \rho&=& \frac{\Pe}{d(d-1)}\nabla^2 \rho+\nabla \cdot (\langle \hat \rho \nabla \hmuint \rangle) \nonumber \\
    &+&\eps \Bigg[\frac{\Pe^2 (1-4d)}{d^3(d-1)^3(d+2)}\nabla^4\rho-\frac{\Pe}{d (d-1)^2} \nabla \cdot \nabla^2 (\langle \hat \rho \nabla \hmuint \rangle)\Bigg]\nonumber
    \\
    &-&\frac{\sqrt{\Pe\eps}}{d-1} \nabla \otimes \nabla : \langle \hat{\mathbf{m}} \otimes \nabla \hmuint\rangle  +O(\eps^{2})\;.\label{eq:oldguys}
\end{eqnarray}
Returning to the original units finally gives
\begin{eqnarray}
    \partial_t \rho&=& D_a\nabla^2 \rho+\nabla \cdot (\langle {\hat \rho} \mu\nabla \hmuint \rangle) \nonumber \\
    &-&\frac{\tilde \gamma}{d} \ell_a^2 D_a \nabla^4 \rho -\frac{\ell_a^2}{d} \nabla \cdot \nabla^2 (\langle \hat \rho \mu\nabla \hmuint \rangle)\nonumber
    \\
    &-&\ell_a \nabla \otimes \nabla : \langle \hat{\mathbf{m}} \otimes \mu\nabla \hmuint\rangle  +O(\ell_a^{4})\;.\label{eq:oldguys}
\end{eqnarray}
Note that $\langle \hat \rho \nabla \hmuint \rangle=-\nabla \cdot {\mathbf \sigma}^{\rm IK}$, where ${\mathbf \sigma}^{\rm IK}$ is the standard Irving-Kirkwood stress tensor~\cite{irvingStatisticalMechanicalTheory1950}. Interestingly, ${\mathbf \sigma}^{\rm IK}$ enters twice in the dynamics: First, as the usual form with a contribution to the current $\nabla \cdot \sigma^{\rm IK}$, and second through a contribution to the current of the form $ \nabla^2  \big(\nabla \cdot \sigma^{\rm IK}\big)$, with a prefactor proportional to $\ell_a^2$. 

All in all, the dynamics can be written in a more familiar form as 
\begin{equation}\label{eq:dynPFAPscorr}
\partial_t\rho=-\mu\partial_i\partial_j \sigma_{ij},\qquad\text{with}\qquad
\sigma_{ij}=-\frac{D_a}\mu\rho \delta_{ij}
-\Big[1-\frac{\ell_a^2}{d}\nabla^2\Big]\sigma^{\rm IK}_{ij}
+\frac{\tilde\gamma D_a}{d \mu}\,\ell_a^2 (\partial_i\partial_j\rho)
+ \sigma^A_{ij},
\end{equation}
and $\sigma^A_{ij}\equiv \ell_a\langle \hat m_i  \partial_j \hmuint \rangle$ is of order $O(\ell_a^2)$.

\if
$\bfJ=\nabla \cdot \sigma$ and the effective stress tensor reads
\begin{equation}
    \sigma = \sigma_{\rm IK}+...
\end{equation}
At this stage, Eq.~\eqref{eq:oldguys} solely involves an expansion in the particle persistence time. 
As such, it can be used to characterize analytically the departure from the passive $\tau\to 0$ limit, which has attracted a lot of interest in the literature~\cite{paliwal2017non}. 
To make progress, one need to close the equation and characterize the correlators $\langle \hat\rho \mu\nabla \hmuint \rangle$ and $\langle \hat{\mathbf{m}} \mu\nabla \hmuint \rangle$. 
This can be done using equations of state that can be measured numerically~\cite{fily_athermal_2012,solon_pressure_2015-1,solon_generalized_2018,omar_mechanical_2023}. 
Alternatively, one can resort to a mean-field approximation, using $\langle \hat\rho \mu\nabla \hmuint \rangle=\rho \mu\nabla \langle \hmuint \rangle$ and $\langle \hat{\mathbf{m}} \mu\nabla \hmuint \rangle={\mathbf{m}} \mu\nabla \langle\hmuint \rangle$. 
The result of this mean-field treatment can actually be obtained directly from the derivation in the presence of an external potential, as detailed below.
\fi
\subsection{Mean-field treatment}
To get a closed expression for the dynamics, one then needs to express the various correlators entering Eq.~\eqref{eq:dynPFAPscorr} in terms of the field $\rho$ and its derivatives. 
This is a standard challenge to describe the dynamics of active particles interacting via pairwise forces, which has attracted much effort in the past~\cite{fily_athermal_2012,solon_pressure_2015-1,solon_generalized_2018,speck2021coexistence,omar2023mechanical}.
 
Here, we instead proceed within a simple mean-field approximation, where $\langle \hat \rho(\br',t)\hat\psi(\br,\bu,t)\rangle\simeq  \rho(\br',t)\psi(\br,\bu,t)$. The result of the previous section can then be recovered and complemented in a few lines starting from Eq.~\eqref{eq:dynpsi}, and using the results of the external-potential Sec.~\ref{sec:derivation}.

Within the mean-field approximation, Eq.~\eqref{eq:dynpsi} indeed becomes
\begin{equation}\label{eq:interactionMF}
    \partial_t\psi+\nabla\cdot\big[(v_a\mathbf u-\mu\nabla \muint)\psi\big]= \frac{1}{\tau_r}\nabla_{u}^2\psi \;,
\end{equation}
where we have introduced the mean-field excess chemical potential
\begin{equation}
\muint[\rho](\br)\equiv\frac{\delta F_{\rm int}[\rho]}{\delta\rho(\br)}=\int d\br'\,w(|\br-\br'|)\rho(\br')\;.
\end{equation}

Note that Eq.~\eqref{eq:interactionMF} is exactly the same expression as the dynamics in the presence of an external potential, Eq.~\eqref{eq:Fokker-Planck}, with $V$ replaced by $\muint$. The derivation thus follows exactly the same route, leading to:
\begin{eqnarray}\label{eq:rhodyn}
\partial_t\rho
&=& D_a\nabla^2\rho
+\mu \nabla \cdot(\rho \nabla \muint) \nonumber\\
&-& \ell_a^2 \tilde \gamma \Da \nabla^4\rho
+\frac{\ell_a^2}{d}\mu\nabla\otimes\nabla : \big[\nabla \muint\otimes\nabla\rho\big] -\frac{\mu \ell_a^2}{d}\nabla^2 \big[ \nabla \cdot(\rho \nabla \muint) \big]+ O(\ell_a^4).
\end{eqnarray}

Equation~\eqref{eq:rhodyn} is exactly what one obtains using Eq.~\eqref{eq:mPFAPs} for $\bm$ and taking the mean-field approximation in Eq.~\eqref{eq:oldguys}. It can be written in the form of a conservation law:
\begin{eqnarray}\label{eq:rhodyn2}
\partial_t\rho&=&-\nabla\cdot J\\
J_i&=& -D_a[1-\tilde\gamma \ell_a^2 \nabla^2]\partial_i\rho
-\mu [1-\frac{\ell_a^2}{d} \nabla^2](\rho \partial_i  \muint) -\frac{\ell_a^2}{d}\mu\partial_j  \big[\partial_i \rho \otimes \partial_j \muint\big] 
 + O(\ell_a^4).
\end{eqnarray}

To proceed further, we expand in gradients of the density field, which allows rewriting the excess chemical potential as
\begin{equation}\label{eq:muint}
  \muint[\rho](\br)=\int d\br' w(|\br-\br'|)\rho(\br')=a_0\rho+ a_2 \nabla^2 \rho  \;,
\end{equation}
where, using the isotropy of $w$, 
\begin{equation}
    a_0 =\int d\br w(|\br|)\qquad \text{and}\qquad a_2 = \frac 1 {2 d} \int d\br w(|\br|) \br^2\;.
\end{equation}
Note that, in Fourier space, the expansion~\eqref{eq:muint} is valid as long as $\mu a_2 k^2 \ll \mu a_0$. Since we expect $a_2 \simeq a_0 \ell_V^2$, where $\ell_V$ is the potential range, it describes the system on length scales which satisfy $k^2 \ll \ell_V^{-2}$. 

Inserting Eq.~\eqref{eq:muint} into the various terms entering Eq.~\eqref{eq:rhodyn} and keeping terms up to fourth order in gradient then leads to
\begin{eqnarray}
   \mu \nabla \cdot(\rho  \nabla \muint) &=&\mu a_0 \nabla \cdot \big(\rho \nabla \rho \big)+\mu a_2\nabla \cdot \big(\rho \nabla \nabla^2 \rho \big)  \;, \\
    \frac{\ell_a^2}{d}\mu\nabla\otimes \nabla : \big[(\nabla \muint)\otimes\nabla\rho\big]&=&\frac{\mu\ell_a^2}{d} a_0 \Big[\frac 1 2 \nabla^2 [(\nabla\rho)^2] + \nabla \cdot [\nabla^2\rho \nabla\rho]\Big] \\
   \frac{\ell_a^2}{d}\Big[\mu\nabla^2 \big( \nabla \cdot(\rho \nabla \muint[\rho]) \big)\Big] &=&\frac{\ell_a^2}{d}\nabla^2 \Big[\mu a_0 \nabla \cdot \big(\rho \nabla \rho \big)\Big] 
\end{eqnarray}

Collecting everything then gives
\begin{eqnarray}
\partial_t\rho
&=&D_a\nabla^2\rho
+\mu a_0 \nabla \cdot \big(\rho \nabla \rho \big)+\mu a_2\nabla \cdot \big(\rho \nabla \nabla^2 \rho \big) \nonumber\\
&-&\ell_a^2\tilde \gamma D_a\nabla^4\rho
+\frac{\mu a_0\ell_a^2}{d}  \Big[\frac 1 2 \nabla^2 [(\nabla\rho)^2] + \nabla \cdot [\nabla^2\rho \nabla\rho]\Big] \nonumber \\
&-&\frac{\mu a_0 \ell_a^2}{d}\nabla^2 \Big[ \nabla \cdot \big(\rho \nabla \rho \big)\Big]+ O(\ell_a^4,\nabla^6).\label{eq:PFAPs final}
\end{eqnarray}

We note that the terms entering Eq.~\eqref{eq:PFAPs final} share a lot of similarity with those found in phenomenological active field theories like active model B and B$^+$~\cite{cates2025active}. 
Of course, since Eq.~\eqref{eq:PFAPs final} is derived from a microscopic model, and since it is not expanded close to a critical point, it has a much more complex structure, including a density-dependent mobility and density-dependent coefficients. While phenomenological field theories are useful to study critical properties, Eq.~\eqref{eq:PFAPs final} reminds us that non-universal ones, like phase diagrams, instead requires properly coarse-grained approaches.

To see how to connect Eq.~\eqref{eq:PFAPs final} to Active Model B$^+$, we first identify the density-dependent mobility $M(\rho)=\rho$. This allows us to rewrite the diffusive term as $D_a\nabla^2\rho=\nabla\cdot[\rho\nabla(D_a\ln\rho)]$. 
The equilibrium-like terms can then be written in terms of a chemical potential. 
Expanding this chemical potential about a critical density, $\rho=\rho_c+\phi$ and imposing the critical-point conditions gives the usual Landau form $\mu_{\rm eq}=r\phi+u\phi^3-K\nabla^2\phi+...$. 

The corrections due to activity generate the two characteristic nonequilibrium operators of Active Model B$^+$, $\nabla^2(\nabla\phi)^2$ and $\nabla\cdot[(\nabla^2\phi)\nabla\phi]$. Using the convention $\partial_t\phi=\nabla^2[\dots+\lambda(\nabla\phi)^2]-\zeta\nabla\cdot[(\nabla^2\phi)\nabla\phi]$, their bare coefficients are $\lambda=\mu a_0\ell_a^2/(2d)$ and $\zeta=-\mu a_0\ell_a^2/d=-2\lambda$. The constraint $\zeta=-2\lambda$ is absent in usual Active Model B$^+$; 
here, it is rooted in the microscopic definition of our particle-based model. We note that our large-scale dynamics is, however, more general than the minimal Active Model B$^+$ theory. In particular, retaining the density dependence of the mobility generates the additional fourth-gradient nonlinearities $\mu a_2\nabla\cdot[\phi\nabla\nabla^2\phi]$ and $-(\mu a_0\ell_a^2/2d)\nabla^4(\phi^2)$ that are absent in Active Model B$^+$.

\section{Non-uniform activity}\label{sec:derivationNONU} 
We now consider the case where the self-propulsion speed $v_a$ depends on the particle position, leading to
\begin{equation}
\partial_t \psi(\mathbf r,\mathbf u,t)
=
-\nabla \cdot \Big[\big[v_a(\br)\mathbf u-\mu\nabla V(\mathbf r)\big] \psi(\mathbf r,\mathbf u,t)\Big]
+\frac{1}{\tau_r}\nabla_{\mathbf u}^2\psi(\mathbf r,\mathbf u,t)\;.
\end{equation}
Rescaling dimensions as before then leads to
\begin{equation}
\partial_t \psi(\mathbf r,\mathbf u,t)
=
-\nabla\cdot\Big[\Big(\sqrt{\frac{\mathrm{Pe}(\br)}{\eps}}\mathbf u
-\nabla V(\mathbf r)\Big)\psi(\mathbf r,\mathbf u,t)\Big]
+\frac{1}{\eps}\nabla_{\mathbf u}^2\psi(\mathbf r,\mathbf u,t)\;, \label{eq:rescaledFPNONU}
\end{equation}
where the P\'eclet number is now position dependent
\begin{equation}
    \Pe(\br) = \frac{d(d-1)\Da(\br)}{v_V \ell_V}\;.
\end{equation}
We still consider a limit where $\tau_r\to 0$ while $D_a(\br)$ remain finite, so that $\mathrm{Pe}\sim O(1)$. For notational simplicity, we introduce
\begin{equation}
    \tilde v_a(\br)=\sqrt{\Pe(\br)}\;.
\end{equation}

Then, the dynamics of $\rho$, $m_i$ and $Q_{ij}$ take the form
\begin{eqnarray}
    \partial_t \rho &=& -\nabla \cdot \Big(
\frac{\tilde v_a}{\sqrt\eps}\mathbf m - \rho \nabla V \Big) \label{eq:denhNONU}\\
\mathcal L_m m_i &=&-\sqrt{\frac{1}{\eps}}\Big(\frac{1}{d}\partial_i (\tilde v_a\rho)+\partial_j (v_a Q_{ij})\Big)\label{eq:maghNONU} \\
\mathcal L_Q Q_{ij}&=&-\sqrt{\frac{1}{\eps}}\left\{\frac{1}{d+2}\partial_k\big(\tilde v_a\delta_{ij} m_k+\tilde v_a\delta_{jk} m_i+\tilde v_a\delta_{ki} m_j\big)-\frac{\delta_{ij}}{d}\partial_k (\tilde v_am_k)+\partial_k (\tilde v_a\mathcal \chi_{ijk})\right\} \;, \label{eq:QhNONU}
\end{eqnarray}
where $\mathcal{\chi}$ is still the third spherical harmonics of $\psi$. The scaling of the fields remains $\rho = O(1)$, $\bm = O(\sqrt{\eps})$, $Q = O(\eps)$, while $\mathcal \chi = O(\eps^{3/2})$ and the inversion of $\mathcal L_m$ and $\mathcal L_Q$ remains unchanged:
\begin{eqnarray}
\mathcal L_m^{-1}
&=&\frac{\eps}{d-1}\left[1-\frac{\eps}{d-1}\big[\partial_t-\partial_j(\partial_j V\,\cdot)\big]\right] + O(\eps^3) \label{eq:Lmm1NONU}\\
\mathcal L_Q^{-1}&=&\frac{\eps}{2d}
+ O(\eps^2)\;.
\end{eqnarray}
Therefore,
\begin{equation}
Q_{ij}=-\frac{\sqrt{\eps}}{2d(d+2)}\left(\partial_i (\tilde v_a \, m_j)+\partial_j (\tilde v_a\,  m_i)-\frac{2}{d}\delta_{ij}\nabla \cdot (\tilde v_a\,  m)\right)+ O(\eps^2).
\end{equation}
Using $\mathcal L_m^{-1}$ to order $\eps$ then gives $m_i=-\frac{\sqrt{\eps}}{d(d-1)}\partial_i (\tilde v_a\,\rho) + O(\eps^{3/2})$, which shows that
\begin{equation}\label{eq:leadingQNONU}
Q_{ij}=\frac{\eps}{d^2(d-1)(d+2)}\left(\partial_i[\tilde v_a \partial_j (\tilde v_a \rho)]-\frac{1}{d}\delta_{ij}\nabla\cdot [\tilde v_a \nabla (\tilde v_a \rho)]\right)+ O(\eps^2).
\end{equation}
Using Eqs.~\eqref{eq:maghNONU},~\eqref{eq:Lmm1NONU}, and~\eqref{eq:leadingQNONU} then gives for $\bm$:
\begin{eqnarray}
m_i&=&-\frac{\sqrt{\eps}}{d(d-1)}\partial_i(\tilde v_a\,\rho) -\frac{\eps^{3/2}}{d(d-1)}
\Bigg[\frac{1}{d^2(d+2)}\partial_i\Big[\tilde v_a \nabla\cdot [\tilde v_a \nabla (\tilde v_a \rho)]\Big]\nonumber\\
&&-\frac{1}{d-1}\Big(\partial_i(\tilde v_a \partial_t\rho)-(\nabla^2 V)\partial_i(\tilde v_a\, \rho)-(\nabla V\!\cdot\!\nabla)\partial_i(\tilde v_a \,\rho)\Big)\Bigg]+ O(\eps^{5/2}).    
\end{eqnarray}

Finally, using this in Eq. \eqref{eq:denhNONU} gives 
\begin{eqnarray}
    \partial_t \rho&=& \frac{1}{d(d-1)}\nabla\cdot [\tilde v_a\, \nabla(\tilde v_a\, \rho)]+\nabla \cdot (\rho \nabla V) 
    +\eps \nabla \cdot \frac{\tilde v_a \nabla \big[\tilde v_a \nabla \cdot \big(\tilde v_a \,\nabla(\tilde v_a \,\rho)\big)\big]}{d^3(d-1)(d+2)}\nonumber \\
    &-& \frac{\eps }{d(d-1)^2}\Big[\nabla \cdot [ \tilde v_a \nabla(\tilde v_a\,\partial_t\rho)]-
    \nabla \cdot [\tilde v_a (\nabla^2 V) \nabla (\tilde v_a \rho)+\tilde v_a (\nabla V \cdot \nabla) \nabla (\tilde v_a \rho)]
    \Big]    
    + O(\eps^{2}) \;.
\end{eqnarray}

Then, inserting the $O(\epsilon^0)$ of $\partial_t \rho$ recursively in the right-hand side yields
\begin{eqnarray}
 \partial_t \rho&=& \frac{1}{d(d-1)}\nabla\cdot [\tilde v_a\, \nabla(\tilde v_a\, \rho)]+\nabla \cdot (\rho \nabla V)\\ 
    &-&\eps  \frac{4d-1}{d^3 (d-1)^3(d+2)}
    \nabla \cdot \Big[\tilde v_a \nabla \Big(\tilde v_a \nabla \cdot \big[\tilde v_a \,\nabla\big(\tilde v_a \,\rho\big)\big]\Big)\Big]
    \nonumber \\
    &-& \frac{\eps}{d(d-1)^2} \nabla \cdot \Big[ 
    \tilde v_a
    \nabla [\tilde v_a\nabla \cdot (\rho \nabla V)]
    - \tilde v_a[(\nabla^2 V) \nabla (\tilde v_a \rho)+ (\nabla V \cdot \nabla) \nabla (\tilde v_a \rho)]\Big]
    + O(\eps^{2}) \;.
\end{eqnarray}
Turning back to the initial variables finally leads to 
\begin{eqnarray}
 \partial_t \rho&=& \frac{\tau_a}{d}\nabla\cdot [v_a\, \nabla(v_a\, \rho)]+\mu \nabla \cdot (\rho \nabla V)\notag\\ 
    &-&\tau_a^3  \frac{4d-1}{d^3 (d+2)}
    \nabla \cdot \Big[v_a \nabla \Big( v_a \nabla \cdot \big[v_a \,\nabla\big(v_a \,\rho\big)\big]\Big)\Big]
    \nonumber \\
    &-&\mu \tau_a^2 \nabla \cdot \Bigg[ 
    \frac{v_a}{d}\Big[
    \nabla [v_a\nabla \cdot (\rho \nabla V)]
    -\nabla^2 V  \nabla(v_a\, \rho)
    -(\nabla V \cdot \nabla) \nabla(v_a\, \rho)]
    \Big]\Bigg]    
    + O(\tau_a^4) \;.\label{eq:vofrfinal}
\end{eqnarray}
We note that the final line involves a term $\propto 
\nabla V 
\otimes \nabla v_a$, which shows that adding confinement along $\hat y$ in the presence of activity modulation along $\hat x$ induces a current along $\hat x$. 
In the absence of the potential, the steady state would be $\psi(\br,\bu)\propto 1/v_a(\br)$ and it would be flux free. 
The steady state of Eq.~\eqref{eq:vofrfinal} already develops a current at order $\tau_a$, but solving the dynamics is required to make this apparent. 
The order $\tau_a^2$ term shows this explicitly.

\section{Conclusion}

Active field theories have proven powerful tools to study the large-scale properties of bulk active fluids. 
However, how to couple them to external potentials or boundaries is an outstanding challenge. This is particularly important since boundaries~\cite{dor2022disordered} and localized inclusions~\cite{baek_generic_2018} have been shown to have long-ranged effects in active fluids.

In this article, we have provided a first step in this direction by developing a systematic perturbative expansion to characterize the dynamics of active particles in the presence of a potential. 
This is the dynamical counterpart to perturbative expansions developed for the steady state of active particles~\cite{fodor_how_2016,woillez2020nonlocal}, which have proven useful in a variety of contexts~\cite{martin_statistical_2021,al2025statistical,al2025statistical2}. 

While we were writing this article, an alternative approach, based on a multi-scale expansion, was proposed to coarse-grain active particles interacting via quorum-sensing~\cite{burekovic2026active}. 
While the two approaches look superficially distinct, we believe that their results should be consistent when applied to the same models.

\vspace{1cm}
\noindent\textbf{Acknowledgments}\\
We thanks Ram Adar for a critical reading of the manuscript and insightful suggestions. This research was supported in part by grant NSF PHY-2309135 to the Kavli Institute for Theoretical Physics (KITP). YK acknowledges financial support from ISF (2038/21), (3457/25) and NSF/BSF (2022605) and thanks Université Paris Cité for a 2026 visiting Professorship.
JT thanks Laboratoire MSC for hospitality. 

\newpage
\appendix{}

\section{Translational diffusivity}
\label{sec:Dp}
Here we repeat the derivation of the coupling to a potential for the case $D_p \neq 0$.
The Fokker-Planck equation is then
\begin{equation}
\partial_t \psi(\mathbf r,\mathbf u,t)
=
-\nabla \cdot \Big[\big[v_a\mathbf u-D_p\nabla -\mu\nabla V(\mathbf r)\big] \psi(\mathbf r,\mathbf u,t)\Big]
+\frac{1}{\tau_r}\nabla_{\mathbf u}^2\psi(\mathbf r,\mathbf u,t)\;.
\end{equation}
Following Sec. \ref{sec:derivation} and rescaling the variables we now find
\begin{equation}
\partial_t \psi(\mathbf r,\mathbf u,t)
=
-\nabla \cdot\Big[\Big(\sqrt{\frac{\mathrm{Pe}}{\eps}}\mathbf u-\tilde D_p\nabla
-\nabla V(\mathbf r)\Big)\psi(\mathbf r,\mathbf u,t)\Big]
+\frac{1}{\eps}\nabla_{\mathbf u}^2\psi(\mathbf r,\mathbf u,t)\;, \label{eq:rescaledFPA}
\end{equation}
where $\tilde D_p=D_p \tau_V/\ell_V^2$ is a passive Péclet number which compares passive diffusion and advection by the potential on a length scale set by the potential variations or its range. For notational consistency, we also drop the tilde on $\tilde \Dp$.

As in the main text it is convenient to introduce the linear operators
\begin{equation}
\mathcal L_m \equiv \partial_t  - D_p \partial_j \partial_j- \partial_j(\partial_j V \cdot)+\frac{d-1}{\eps}\qquad\text{and}\qquad \mathcal L_Q \equiv \partial_t - D_p \partial_j \partial_j-\partial_j(\partial_j V\cdot)+\frac{2d}{\eps}\;.
\end{equation}
Then, the dynamics of $\rho$, $m_i$ and $Q_{ij}$ take the form
\begin{eqnarray}
    \partial_t \rho &=& -\nabla \cdot \Big(
\sqrt{\frac{\Pe}{\eps}}\mathbf m - D_p \nabla\rho- \rho \nabla V \Big) \label{eq:denhA}\\
\mathcal L_m m_i &=&-\sqrt{\frac{\Pe}{\eps}}\Big(\frac{1}{d}\partial_i \rho+\partial_j Q_{ij}\Big)\label{eq:maghA} \\
\mathcal L_Q Q_{ij}&=&-\sqrt{\frac{\Pe}{\eps}}\left\{\frac{1}{d+2}\partial_k\big(\delta_{ij} m_k+\delta_{jk} m_i+\delta_{ki} m_j\big)-\frac{\delta_{ij}}{d}\partial_k m_k+\partial_k \mathcal \chi_{ijk}\right\} \;, \label{eq:QhA}
\end{eqnarray}

To proceed we invert the linear operators $\mathcal L_m$ and $\mathcal L_Q$ as
\begin{eqnarray}
\mathcal L_m^{-1}
&=&\frac{\eps}{d-1}\left[1-\frac{\eps}{d-1}\big(\partial_t- D_p \partial_j \partial_j-\partial_j[\partial_j V\,\cdot]\big)\right] + O(\eps^3) \label{eq:Lmm1A}\\
\mathcal L_Q^{-1}&=&\frac{\eps}{2d}
+ O(\eps^2)\;.
\end{eqnarray}
Following Sec. \ref{sec:derivation} of the main text, we then find for $\bm$:
\begin{equation}
\mathbf m=-\frac{\sqrt{\Pe\,\eps}}{d(d-1)}\nabla\rho -\frac{\eps^{3/2}}{d(d-1)}
\Bigg[\frac{(\Pe)^{3/2}}{d^2(d+2)}\nabla(\nabla^2\rho)-\frac{\sqrt{\Pe}}{d-1}\Big(\nabla\partial_t\rho- D_p \nabla^2 \nabla\rho-(\nabla^2 V)\nabla\rho-(\nabla V\!\cdot\!\nabla)\nabla\rho\Big)\Bigg]+ O(\eps^{5/2}).\label{eq:mwithDp}
\end{equation}
This then leads to
\begin{eqnarray}
    \partial_t \rho 
    &=& \Big( D_p+ \frac{\Pe}{d(d-1)}\Big)\nabla^2 \rho+\nabla \cdot (\rho \nabla V) \nonumber \\
    &+&\eps \Bigg[\Big(\frac{\Pe^2 }{d^3(d-1)(d+2)}\nabla^4\rho+\frac{\Pe}{d (d-1)^2}\nabla \otimes \nabla: \nabla V\otimes\nabla\rho\\
    &-&\frac{\Pe}{d (d-1)^2}\nabla^2(\partial_t \rho-D_p \nabla^2 \rho)\Bigg]\nonumber + O(\eps^{2})
\end{eqnarray}
Finally, inserting the $O(\epsilon^0)$ of $\partial_t \rho$ recursively in the right-hand side yields
\begin{eqnarray}
    \partial_t \rho 
    &=& \Big( D_p+ \frac{\Pe}{d(d-1)}\Big)\nabla^2 \rho+\nabla \cdot (\rho \nabla V) \nonumber \\
    &+&\eps \Bigg[\Big(\frac{\Pe^2 (1-4d)}{d^3(d-1)^3(d+2)}\nabla^4\rho+\frac{\Pe}{d (d-1)^2}\nabla \otimes \nabla: \nabla V\otimes\nabla\rho\\
    &-&\frac{\Pe}{d (d-1)^2}\nabla^2\nabla \cdot (\rho \nabla V)\Bigg]\nonumber + O(\eps^{2})
\end{eqnarray}
Turning back to the initial variables leads to Eq.~\eqref{eq:potfieldtheorytd}. Note that $ D_p$ does not contribute to the coefficient of $\nabla^4\rho$ in the time-dependent equation because the $ D_p$ dependent pieces cancel between the ${\bf m}$ correction and the recursive elimination of $\nabla^2\partial_t\rho$. Similarly, the steady-state equation is directly obtained from Eqs.~\eqref{eq:denhA} and~\eqref{eq:mwithDp}.

\section{Run-and-tumble particles}
\label{sec:RTP}
Run-and-tumble particles in external potentials have also received significant interest~\cite{tailleur_statistical_2008,tailleur2009sedimentation,elgeti2015run,angelani2014first,dhar2019run,santra2020run,de2023run}. 
In this Appendix we construct their hydrodynamic description by adapting the derivation of Sec.~\ref{sec:derivation}, keeping $D_p=0$. The Master equation now reads:
\begin{equation}
\partial_t \psi(\mathbf r,\mathbf u,t)
=
-\nabla \cdot \Big[\big[v_a\mathbf u-\mu\nabla V(\mathbf r)\big] \psi(\mathbf r,\mathbf u,t)\Big] -\frac{1}{\tau_T} \psi(\mathbf r,\mathbf u,t)+\frac{1}{\tau_T} \frac{\rho(\mathbf r,t)}{\Omega_d} \;,
\end{equation}
where $\tau_T^{-1}$ is the tumbling rate. After rescaling the coordinate as detailed in Sec. \ref{sec:derivation} we obtain
\begin{equation}
\partial_t \psi(\mathbf r,\mathbf u,t)
=
-\nabla\cdot\Big[\Big(\sqrt{\frac{\mathrm{Pe}}{\eps}}\mathbf u
-\nabla V(\mathbf r)\Big)\psi(\mathbf r,\mathbf u,t)\Big]
-\frac{1}{\eps} \psi(\mathbf r,\mathbf u,t)+\frac{1}{\eps} \frac{\rho(\mathbf r,t)}{\Omega_d} \;, \label{eq:rescaledFPRTP}
\end{equation}
where we introduced $\eps=\tau_T/\tau_V$ and a P\'eclet number
\begin{equation}
    \Pe = 
\frac{v_a^2 \tau_{T}}{v_V \ell_V}\;.
\end{equation}
Then we use \eqref{eq:rescaledFPRTP} to derive equations for the density, magnetization, and nematic tensor. It is again convenient to introduce the linear operators
\begin{equation}
\mathcal L_m \equiv    \partial_t - \partial_j(\partial_j V \cdot)+\frac{1}{\eps}\qquad\text{and}\qquad \mathcal L_Q \equiv \partial_t -\partial_j(\partial_j V\cdot)+\frac{1}{\eps}\;,
\end{equation}
so that the dynamics of $\rho$, $m_i$ and $Q_{ij}$ take the form
\begin{eqnarray}
    \partial_t \rho &=& -\nabla \cdot \Big(
\sqrt{\frac{\Pe}{\eps}}\mathbf m - \rho \nabla V \Big) \label{eq:denhRTP}\\
\mathcal L_m m_i &=&-\sqrt{\frac{\Pe}{\eps}}\Big(\frac{1}{d}\partial_i \rho+\partial_j Q_{ij}\Big)\label{eq:maghRTP} \\
\mathcal L_Q Q_{ij}&=&-\sqrt{\frac{\Pe}{\eps}}\left\{\frac{1}{d+2}\partial_k\big(\delta_{ij} m_k+\delta_{jk} m_i+\delta_{ki} m_j\big)-\frac{\delta_{ij}}{d}\partial_k m_k+\partial_k \mathcal \chi_{ijk}\right\} \;, \label{eq:QhRTP}
\end{eqnarray}
Inverting the linear operators $\mathcal L_m$ and $\mathcal L_Q$ as
\begin{eqnarray}
\mathcal L_m^{-1}
&=&\eps \left[1-\eps\big(\partial_t-\partial_j(\partial_j V\,\cdot)\big)\right] + O(\eps^3) \label{eq:Lmm1RTP}\\
\mathcal L_Q^{-1}&=&\eps
+ O(\eps^2)\;.
\end{eqnarray}
Therefore,
\begin{equation}
Q_{ij}=-\frac{\sqrt{\Pe\,\eps}}{(d+2)}\left(\partial_i m_j+\partial_j m_i-\frac{2}{d}\delta_{ij}\nabla \cdot m\right)+ O(\eps^2).
\end{equation}
Using $\mathcal L_m^{-1}$ to order $\eps$ then gives $m_i=-\frac{\sqrt{\Pe\,\eps}}{d}\partial_i \rho + O(\eps^{3/2})$, which shows that
\begin{equation}\label{eq:leadingQRTP}
Q_{ij}=\frac{2\Pe \, \eps}{d(d+2)}\left(\partial_i\partial_j \rho-\frac{1}{d}\delta_{ij}\nabla^2\rho\right)+ O(\eps^2).
\end{equation}
Using these then gives 
\begin{equation}
\mathbf m=-\frac{\sqrt{\Pe\,\eps}}{d}\nabla\rho -\frac{\eps^{3/2}}{d}
\Bigg[\frac{2(d-1) (\Pe)^{3/2}}{d(d+2)}\nabla(\nabla^2\rho)-\sqrt{\Pe}\Big(\nabla\partial_t\rho-(\nabla^2 V)\nabla\rho-(\nabla V \cdot \nabla)\nabla\rho\Big)\Bigg]+ O(\eps^{5/2}).
\end{equation}
Finally, using this in Eq.~\eqref{eq:denhRTP} gives 
\begin{eqnarray}
    \partial_t \rho&=& \frac{\Pe}{d} \nabla^2 \rho+\nabla \cdot (\rho \nabla V) \nonumber \\
    &+&\eps \Bigg[\frac{2(d-1)\Pe^2}{d^2(d+2)}(\nabla^4\rho)-\frac{\Pe}{d}\Big(\nabla^2\partial_t\rho-\nabla \otimes \nabla: \nabla V\otimes\nabla\rho\Big)\Bigg]+ O(\eps^{2}) \;.
\end{eqnarray}
Inserting the $O(\epsilon^0)$ of $\partial_t \rho$ recursively yields
\begin{eqnarray}
\partial_t\rho &=& \frac{\Pe}{d} \nabla^2 \rho +\nabla\cdot(\rho\nabla V) \nonumber \\
&+&\frac{\Pe \, \eps}{d} \Big[\frac{\Pe(d-4)}{d(d+2)} \nabla^4\rho +
\nabla \otimes \nabla:\nabla V \otimes \nabla\rho- \nabla^2\nabla\cdot(\rho\nabla V)\Big]
+O(\varepsilon^2)\,.
\end{eqnarray}
Finally, defining $D_a=v_a^2 \tau_T/d$ and $\ell_a=v_a \tau_T$, the equation can be written as:
\begin{eqnarray}
\partial_t\rho
&=& D_a \nabla^2\rho
+\mu \nabla\cdot(\rho\nabla V) \nonumber\\
&+&\frac{d-4}{d+2}\frac{\ell_a^2}{d} D_a \nabla^4\rho
+\frac{\mu \ell_a^2}{d} \nabla \otimes \nabla:\nabla V \otimes \nabla\rho-\frac{\mu \ell_a^2}{d} \nabla^2\nabla\cdot(\rho\nabla V)+ O(\ell_a^4)\;.
\end{eqnarray}
We note that, unlike for ABPs, the $O(\nabla^4)$ predicts a linear instability at large $q$ in $d>4$. However, this would happen on scales much smaller than $\ell_a$, which are not described by this perturbative approach.

\printbibliography

\end{document}